\documentclass{aa}
\usepackage{graphicx}
\usepackage{rotating}
\usepackage{placeins}
\begin{document}

\title{Tidal interaction vs. ram pressure stripping effects as seen\\ in X-rays$^*$}
\subtitle{Hot gas in group and cluster galaxies}

\author{ 
M. We\.zgowiec\inst{1,3}
 \and D.J. Bomans\inst{1}
 \and M. Ehle\inst{2}
 \and K.T. Chy\.zy\inst{3}
 \and M. Urbanik\inst{3}
 \and J. Braine\inst{4}
 \and M. Soida\inst{3}}
\institute{Astronomisches Institut der Ruhr-Universit\"at Bochum, Universit\"atsstrasse 150, 44780 Bochum, Germany
\and ESAC, XMM-Newton Science Operations Centre, P.O. Box 78, 28691 Villanueva de la Ca\~nada, Madrid, Spain
\and Obserwatorium Astronomiczne Uniwersytetu Jagiello\'nskiego, ul. Orla 171, 30-244 Krak\'ow, Poland
\and Laboratoire d'Astrophysique de Bordeaux (OASU), Universite de Bordeaux, CNRS/INSU, BP 89, 33270 Floirac, France}

\offprints{M. We\.zgowiec}
\mail{mawez@astro.rub.de\\
$^{*}$Based on observations obtained with XMM-Newton, an ESA science mission with instruments and contributions
directly funded by ESA Member States and NASA.}
\date{Received date/Accepted date}

\titlerunning{Hot gas in group and cluster galaxies}
\authorrunning{M. We\.zgowiec et al.}

\abstract
%context
{
Cluster and group spiral galaxies are very often affected by their environment. 
The hot intracluster/intragroup medium (ICM/IGM) and a high galaxy density can lead to perturbations of the galactic interstellar medium (ISM) due 
to ram pressure and/or tidal interaction effects. In radio polarimetry 
observations, both phenomena may manifest similar features. X-ray data can help to determine the real origin of the perturbation.
}
%aims 
{
We analyse the distribution and physical properties of the hot gas in the Virgo cluster spiral galaxies NGC\,4254 and NGC\,4569, which indicate that 
the cluster environment has had a significant influence on their properties.
By performing both spatial and spectral analyses of X-ray data, we try to distinguish between two major phenomena: tidal and ram pressure interactions. 
We compare our findings with the case of NGC\,2276, in which a shock was reported, by analysing XMM-Newton X-ray data for this galaxy. 
}
%methods
{
We use archival XMM-Newton observations of NGC\,4254, NGC\,4569, and NGC\,2276. 
Maps of the soft diffuse emission in the energy band 0.2 - 1 keV are obtained.
For the three galaxies, especially at the position of magnetic field enhancements we perform a spectral analysis to derive gas temperatures and thus to look for
shock signatures. A shock is a signature of ram pressure resulting from supersonic velocities; weak tidal interactions are not expected to influence the temperature of the ionized gas.
}
%results
{
In NGC\,4254, we do not observe any temperature increase at the position of the bright polarized radio ridge. This suggests that the feature is formed by tidal
interactions, and not by ram pressure stripping. 
NGC\,4569 shows a higher temperature at the position of the polarized features, which may be indicative of ram-pressure effects. 
For NGC\,2276, we do not find clear indications of a shock. Although ram-pressure effects 
seem to be visible, the main driver of the observed distortions is most likely tidal interaction.  
} 
%conclusions 
{
Determining gas temperatures via sensitive X-ray observations at the position of polarized radio ridges seems to be a good 
method for distinguishing between ram pressure and tidal interaction effects acting upon a galaxy.
}

\maketitle

\section{Introduction}
\label{intro}
  
Cluster and group galaxies experience a variety of environmental effects, which result in \ion{H}{i} deficiences, the truncations of their \ion{H}{i} disks, 
or the triggering of star formation, as well as cause changes 
in the global morphologies of galaxies (e.g. Boselli \& Gavazzi~\cite{bosgav}). The interstellar medium (ISM) of disk galaxies can undergo significant transformations, 
particularly gas loss (e.g. Vollmer et al.~\cite{vollmer01}), either due to ram pressure, where a galaxy is moving through the group/cluster medium, or tidal effects. 
Interesting features caused by these effects are the polarized radio ridges often seen in perturbed galaxies (e.g. Chy\.zy et al.~\cite{chyzy4254}, 
Vollmer et al.~\cite{letter}). Radio observations alone are unable to distinguish unambiguously whether the observed polarized radio ridge was produced by tidal or ram pressure effects. 
In both cases, the magnetic field is locally enhanced and ordered, but the mechanisms causing that are different. Ram pressure effects create compressions of the magnetic field and gas, while 
tidal interactions cause stretching and shearing forces that act upon magnetic fields, leading to the production of an anisotropic component and an enhancement of the polarized radio emission 
(Chy\.zy~\cite{chyzy4254b}). With the use of X-ray observations of the hot gas in galaxies, one can obtain evidence in favour of one of these mechanisms.
Examining the hot gas in a cluster galaxy provides an opportunity to study the ISM, as well as its 
interactions with the intracluster/intragroup medium (ICM/IGM), and insight into the history of a galaxy (e.g. Machacek et al.~\cite{machacek}). To determine temperatures and 
densities of the ISM, high-quality sensitive X-ray data are necessary for both spatial and spectral analyses. 

When investigating radio polarized ridges in the X-ray domain, we look for local increases in the temperature of the hot gas at the galaxy-ICM/IGM interface. 
If the temperature is higher than that of the surrounding gas, we can expect shock-heating. This however occurs only if the galaxy moves at supersonic velocities. To calculate the local speed of sound, 
we use the formula (e.g. Lequeux~\cite{lequeux05})

\begin{equation}
c_s = \sqrt{\frac{{\gamma}kT}{{\mu}m_H}},
\end{equation}
where $k$ is the Boltzmann constant, T is the temperature of the gas, $\mu$ is the molecular weight (for fully ionized plasma we adopt a value of 0.7, following Lequeux~\cite{lequeux05}),
$\gamma$ is the adiabatic index and, $m_H$ is the mass of a hydrogen atom.
%Since in various papers the sound speed value is calculated with varying accuracy, we perform our calculations without any simplifications of intermediate results, in order to obtain a precise
%value of the sound speed. This is very important, since we do not expect large speed differences, hence low Mach numbers. In such situation, an uncertainty of the order of 5\% can already be
%critical. 
In the case of a fully ionized plasma (${\gamma}=5/3$), the formula simplifies to $c_s = 0.14\sqrt{T}$, where the sound speed $c_s$ is in km/s and temperature $T$ in Kelvins.
For the relation between gas temperatures and the Mach number of a shock, we use Eq. 2 from Rasmussen et al.~(\cite{rasmus})

\begin{equation}
\frac{T_2}{T_1} = \frac{(1-{\gamma}+2{\gamma}M^2)({\gamma}-1+2/M^2)}{(1+{\gamma})^2}.
\end{equation}
This formula can be well-approximated by ${T_2}/{T_1}\simeq M$, where $T_2$ and $T_1$ are post- and
preshock temperatures of the hot gas, respectively, and $M$ is the Mach number of the shock. 
Particularly interesting areas to search for the temperature increases that could be produced by shocks in the plasma are regions of radio polarized ridges, which are caused by strong
enhancements of the magnetic field. This enhancement of the magnetic field can be caused by the compression resulting from the supersonic speed of a galaxy. When a compression is produced by
a shock, we see an increase in the temperature of the hot gas. Enhancements of the magnetic fields can also be produced however by shearing forces, which are associated with tidal
interactions and should not generate an increase in temperature. Tidally induced shock heating is also known to exist, but this occurs instead during the merging of galaxies 
(e.g. Sinha \& Holley-Bockelmann~\cite{sinbock}). Such a process is much more violent than close encounters of galaxies that have a high mass ratio, as discussed in this paper.  
Therefore, measuring temperatures in regions of radio polarized ridges provides a possibility of distinguishing between the tidal or ram-pressure origin
of the observed enhancement of the magnetic field. One needs, however, to be aware that a local enhancement of the star formation can also lead to an increase in the temperature of the hot gas. 
Nevertheless, radio polarized ridges are often observed in the outskirts of galactic disks (i.e. at the galaxy-ICM/IGM interface), where agglomerations of young stars are not observed.

We present our analysis of X-ray data for two Virgo cluster\footnote{We assume that the distance to the Virgo cluster is 17\,Mpc.}
spiral galaxies NGC\,4254 and NGC\,4569, as well as NGC\,2276 in the NGC\,2300 group of galaxies. 

\begin{figure}[ht]
\resizebox{\hsize}{!}{\includegraphics{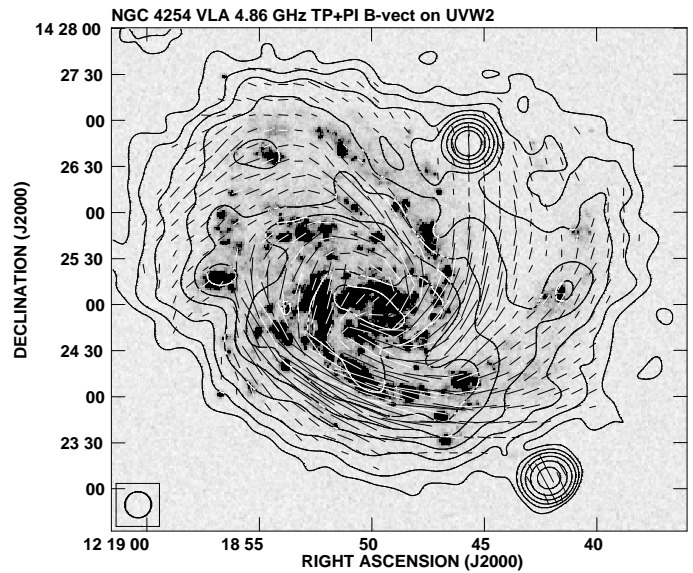}}
\caption{
                Contours of the total power radio continuum of NGC\,4254. The radio image at 4.86\,GHz with a resolution of $15\arcsec$ and $\vec{B}$--vectors proportional to the polarized emission
                (Chy\.zy et al.~\cite{chyzy4254}) is overlaid
                on the XMM-Newton Optical Monitor UVW2 filter image. The contours are at: 3, 8, 16, 32, 64, 128, 230, and 350 $\times$ 10 $\mu$Jy/beam area. Vectors of length of $1\arcsec$ correspond to
                a polarized intensity of 10$\mu$Jy/beam area. The beam size is shown in the lower left corner of the figure.
                }
\label{4254radio}
\end{figure}

NGC\,4254 is an Sc galaxy situated outside the Virgo cluster X-ray cloud (B\"ohringer et al.~\cite{bohringer}),
at an angular distance of $3\fdg 56$ from the centre (1.07\,Mpc in the plane of the sky).
It is a rapidly star-forming
galaxy with no significant signs of an \ion{H}{i} deficiency (Cayatte et al.~\cite{cayatte94}). 
An asymmetric distribution of the radio emission is accompanied by a bright polarized ridge with a parallel magnetic field
at the southern edge of the disk, outside the bright star-forming arm (see Fig.~\ref{4254radio}).
An XMM-Newton map of the soft X-ray emission of NGC\,4254 was previously published in Chy\.zy et al.~(\cite{chyzy4254}),
where sensitive high resolution VLA radio data were also presented. 

\begin{figure}[ht]
\resizebox{\hsize}{!}{\includegraphics{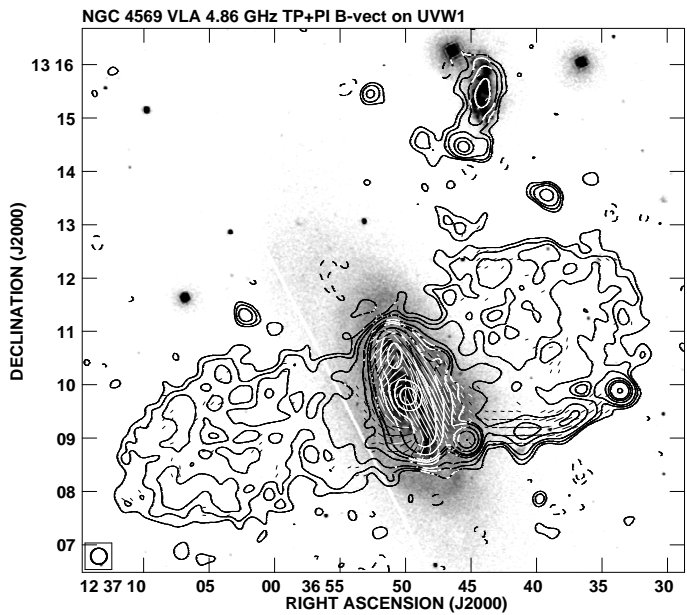}}
                \caption{
                Total power radio continuum map of NGC\,4569. The radio image at 4.86~GHz with superimposed B-vectors of polarized intensity (Chy\.zy et al. in prep.),
                is overlaid on the XMM-Newton Optical Monitor UVW1 filter image.
                The contours are -5, -3, 3, 5, 8, 16, 24, 32, 64, 128, 256, 512, and 1024 $\times$ 9~$\mu$Jy/beam area.
                The polarization vector of 1$'$ corresponds to a polarized intensity of 0.4~mJy/beam area. The angular resolution is 20$"$. The beam size is shown in the lower left corner
                of the figure.
                }
                \label{4569radio}
        \end{figure}

NGC\,4569 is an SABa galaxy situated relatively close to the Virgo cluster centre, at a distance of $1\fdg 66$ (500\,kpc in the plane of the sky)
from Virgo A. It is an anaemic galaxy showing extended radio lobes unusual for
a normal spiral, which can be attributed to previous intense star-formation activity, that has now significantly diminished (Chy\.zy et al.~\cite{chyzy4569}). 
At the southern edge of the western lobe, we can see a ridge of polarized emission that is parallel to the magnetic field (see Fig.~\ref{4569radio}).
For a detailed presentation and discussion of the radio data from this galaxy, we refer to Chy\.zy et al.~(\cite{chyzy4569} and in prep.).
The large-scale distribution of the hot X-ray gas of NGC\,4569 was presented in a paper discussing observational evidence of Mach cones around Virgo cluster galaxies (We\.zgowiec et 
al.~\cite{wezgowiec11}).

Analyses of radio data for NGC\,4254 (Chy\.zy et al.~\cite{chyzy4254} and Chy\.zy~\cite{chyzy4254b}) and NGC\,4569 
(Chy\.zy et al.~in prep.) have uncovered in both galaxies extended, strongly polarized ridges. The authors argue that these ridges could be caused by the interaction 
with the cluster medium in the case of NGC\,4569 and tidal interaction with a companion in the case of NGC\,4254. The radio polarimetry observations alone, however, cannot state 
unambiguously which of the two mechanisms is at work. To investigate these cases in greater detail, we performed sensitive X-ray observations of both galaxies 
with the XMM-Newton Space Telescope (Jansen et al.~\cite{jansen}). 

We also make use of archival XMM-Newton observations of NGC\,2276 in the NGC\,2300 galaxy group of galaxies. 

NGC\,2276 is an SABc galaxy with a lopsided optical structure and a slight extension towards NGC\,2300 at the centre of the galaxy group. One of the spiral arms
seems to follow this extension. Star forming regions are somewhat shifted to the western and southwestern side of the galaxy, where a distinct and sharp edge is visible.
Radio observations by Hummel \& Beck~(\cite{humbeck}) showed a steeper gradient in the total radio emission and an associated polarized radio ridge in this part of the galaxy, with 
the magnetic field parallel to the ridge,
as well as a slightly polarized radio tail extending to the east on the other side of the galaxy.
For this galaxy, data from the Chandra X-ray Observatory was presented by Rasmussen et al.~(\cite{rasmus}). The authors argued that NGC\,2276 was displaying 
a shock front along the western edge of the disk and that this most likely could be a result of a rapid passage through the hot IGM. 

Therefore, this galaxy is a good case to compare with the two Virgo cluster galaxies. 
We also compare our analysis of XMM-Newton observations of extended emission from NGC\,2276 to the results of Rasmussen et al.~(\cite{rasmus}).   

\section{Observations and data reduction}
\label{obsred}

\begin{table*}[ht]
        \caption{\label{objects}Basic astronomical properties of studied galaxies}
\centering
                \begin{tabular}{cccccccc}
\hline\hline
\vspace{1pt} NGC &\vspace{1pt} Morph. &\vspace{-9pt} Optical position\tablefootmark{a} &\vspace{1pt} Incl.\tablefootmark{a} &\vspace{1pt} Pos. & Proj. dist. to & Radial & log(L$_B$/L$_{\sun}$)\tablefootmark{d}  \\
\vspace{5pt} & type\tablefootmark{a}& \hspace{5pt} $\textstyle\alpha_{2000}$\hspace{30pt}$\textstyle\delta_{2000}$ & [\degr]& ang.\tablefootmark{a} & system centre\tablefootmark{b} & velocity\tablefootmark{c} & \\
& & &[\degr]&[\degr] &[\degr]/[Mpc] & [km/s] & \\
\hline
\vspace{5pt}
2276 & SABc & 7$^{\rm h}$27$^{\rm m}$14\fs3 \hspace{1pt} +85\degr 45\arcmin 16\arcsec & 40 & 110 & 0.1 & 240 & 10.67 \\
\vspace{5pt}
4254 & Sc & 12$^{\rm h}$18$^{\rm m}$49\fs6 \hspace{1pt} +14\degr 24\arcmin 59\arcsec & 32 & 60 & 3.56/1.07 & 1130 & 10.53 \\
\vspace{5pt}
4569 & SABa & 12$^{\rm h}$36$^{\rm m}$50\fs1 \hspace{1pt} +13\degr 09\arcmin 46\arcsec & 66 & 23 & 1.66/0.5 & 1530 &  10.70 \\
\hline
\end{tabular}
\tablefoot{
\tablefoottext{a}{taken from HYPERLEDA database -- http://leda.univ-lyon1.fr -- see Paturel et al.~(\cite{leda}).}
\tablefoottext{b}{for NGC\,2276 a projected distance to the most massive galaxy NGC\,2300 is provided.}
\tablefoottext{c}{relative to the IGM/ICM estimated using the redshift based velocities in the HYPERLEDA database.}
\tablefoottext{d}{Intrinsic blue luminosities in solar luminosities -- Tully~\cite{tully88}.}
}
\end{table*}

Basic astronomical properties of the galaxies presented in this paper are shown in Table~\ref{objects}.
The observations of Virgo cluster spiral galaxies NGC\,4254 and NGC\,4569 using XMM-Newton were performed on
29 June 2003 and 13/14 December 2004, respectively. The observations of the NGC\,2300 group were performed on 16 March 2001 (see Table~\ref{xobs}).

To fully meet the need for sensitive observations that would allow the detailed investigation of diffuse soft X-ray emission, long
observations using the thin filter for the EPIC pn camera were performed (see Table~\ref{xobs}). 
The data were processed using the SAS 9.0 package (Gabriel et al.~\cite{sas})
with standard reduction procedures. Following the routine of tasks $emchain$ and $epchain$, we obtained event lists 
for the two MOS cameras (Turner et al.~\cite{turner}) and the pn camera (Str\"uder et al.~\cite{strueder})
for each galaxy. Next, the event lists were then filtered for 
periods of intense radiation of high energy background, using the light curves extracted in the 10-12\,keV energy band from the pn data and above 10\,keV from the MOS data:
in the case of the observations of NGC\,4569, low external background
provided high quality data throughout the whole observation. For the observations
of the NGC2300 group, the same good observational
conditions were met.
In the case of NGC\,4254, long periods of high flaring background resulted in a significant
reduction in the observation time useful for our purposes. However, the data are still sufficient for the spectral
analysis owing to the high X-ray luminosity of this galaxy, which provided enough counts. 

The resulting event lists were checked again, this time for the residual exsistence of soft proton-flare contamination, which
could influence the faint extended emission. To do that, we used a script\footnote{http://xmm2.esac.esa.int/external/xmm\_sw\_cal/\\background/epic\_scripts.shtml\#flare}
that performs calculations developed by Molendi et al.~(\cite{spcheck}).
We found out that the event lists remained contaminated to some extent in the case of NGC\,4254 and NGC\,4569. 
The possible influence of this contamination is discussed in Sect.~\ref{spectra} in more detail.

The filtered event lists were used to produce images, background images, and exposure maps, 
which were masked for an acceptable detector area (used for imaging and source detection, i.e. without gaps between CCDs and either noisy rows or columns) 
with a vignetting correction using the {\it images}
script\footnote{http://xmm.esac.esa.int/external/xmm\_science/\\gallery/utils/images.shtml}.
All images and maps were produced in the bands of 0.2 - 1 keV, 1 - 2 keV, 2 - 4.5 keV, and 4.5 - 12 keV. 
Images for all bands were then used to search for background point sources using the standard SAS
{\it edetect\_chain} procedure. Regions found to include a possible point-like source were then excluded from the event lists. The area was individually chosen 
by eye for each source, to ensure exclusion of all pixels brighter than the surrounding background. The whole procedure 
of creating images in the same bands was then repeated. 
The final point-source-free images were smoothed with a Gaussian filter to obtain a resolution of 10$\arcsec$ FWHM and combined to create a total EPIC image for each energy band.
Since this paper focuses on the soft extended emission, a high signal-to-noise ratio and thus better sensitivity to extended structures is needed. 
We achieved this, as well as removed any artifacts introduced by the combination of the EPIC images, which resulted from either an imperfect exposure correction, especially close to the gaps, or 
excluded bad rows, by again smoothing the final images with a Gaussian filter, 
using the AIPS package, to achieve a final resolution of 30$\arcsec$ FWHM. 
Although smoothing over the ''source holes'' with a large Gaussian filter introduces artificial depressions in the surface-brightness 
to the images, we confirmed that this did not produce any artificial extensions, as might have been the case when the point sources were blended with the extended emission. 
Furthermore, no science data were extracted from the images directly, as their main purpose was to show the distribution of the low surface-brightness extended emission. 
%Last, but not least, the final smoothing was 
%performed using a smoothing scale much larger than the areas for the point-source exclusion, which makes the artificial depressions negligible. 
This smoothing of the images was also done to allow a better comparison with the radio maps. 
The noise values given by the root mean square (rms) in the final maps were obtained by averaging the emission over a large source-free area. 
Those values were used for the construction of the contour maps of the X-ray emission.

The spectral analysis was then performed. 
To create spectra, only the point-source-free event list from the pn camera was used as it offers the highest sensitivity in the soft energy band.
For selected regions, spectra were acquired. Similarly, the spectra of the background were obtained using blank-sky 
event lists (see Carter \& Read~\cite{carter}). The whole-sky event lists were selected, as the ''tailored'' blank-sky event lists (e.g. in the direction to the source, hence 
with similar column density) are of inferior statistical quality for NGC\,4254 and NGC\,4569. 

The blank-sky event lists were filtered using the same procedures as for the source event lists.
For each spectrum, both response matrices and effective area files were produced. For the latter, detector maps needed for the extended emission analysis
were also created. Next, the spectra were binned, which resulted in a higher signal-to-noise ratio. To get a reasonable number of bins at the same time, we chose 
to have 25 total counts in one energy bin. The binning method took into account the energy resolution, which is energy dependent. 
Finally, the spectra were fitted using XSPEC~11 (Arnaud~\cite{xspec}).

For NGC\,2276, we followed the analysis of Rasmussen et al.~\cite{rasmus} and excluded the softest emission
for the spectral fitting of the IGM in order to suppress the influence of the high Galactic foreground absorption in the direction towards this source, which resulted in lower soft X-ray 
emission from the source and consequently might have led to an oversubtraction of the blank-sky background spectra.

For NGC\,4254 and NGC\,4569, we used the Optical Monitor data acquired during the same observations and produced images in UVW1 and UVW2 filters using the standard SAS {\it omichain} procedure, 
to create overlays of the radio continuum data, presented in Figs.~\ref{4254radio} and \ref{4569radio}.

\begin{table*}[ht]
\caption{\label{xobs}Parameters of X-ray observations of studied galaxies}
\centering
\begin{tabular}{lllllllll}
\hline\hline
NGC & Obs. ID & Obs. date & Exp. time\tablefootmark{a} & pn filter & pn obs. & MOS filter & MOS obs. & nH\tablefootmark{c} \\
\vspace{5pt} & & & & & mode\tablefootmark{b} & & mode\tablefootmark{b} & \\
\hline
\vspace{5pt}
2276 & 0022340201 & 2001-03-16 & 54.8 (48.2) & Thin & FF & Thin & FF & 5.69\\
\vspace{5pt}
4254 & 0147610101 & 2003-06-29 & 43.2 (12.8) & Thin & EF & Thin & FF & 2.81\\
\vspace{5pt}
4569 & 0200650101 & 2004-12-13/14 & 66 (49) & Thin & EF & Medium & FF & 2.82\\
\hline
\end{tabular}
\tablefoot{
\tablefoottext{a}{Total time in ksec with clean time for pn camera in parentheses.}
\tablefoottext{b}{Observing mode: FF - Full frame, EF - Extended full frame.}
\tablefoottext{c}{Column density in [10$^{20}$ cm$^{-2}$] weighted average value after LAB Survey of Galactic \ion{H}{i}, see Kalberla et al.~(\cite{lab}).}
}
\end{table*}

\section{Results}
\label{results}

\subsection{NGC\,4254}
%\label{4254r}

%NGC\,4254 is an Sc galaxy situated outside the Virgo Cluster X-ray cloud (B\"ohringer et al.~\cite{bohringer}),
%at a distance of $3\fdg 56$ from the centre (1.07\,Mpc in the sky plane).
%It is a rapidly star-forming
%galaxy with no signs of any gas deficiency. An asymmetric distribution of the radio emission is accompanied by a bright polarized ridge with parallel magnetic field
%at the southern gde of the disk (see Fig.~\ref{4254radio}).
%A thorough analysis of the radio data and of magnetic fields can be found in
%Soida et al.~(\cite{soida96}), as well as in Chy\.zy et al.~(\cite{chyzy4254}) and Chy\.zy~(\cite{chyzy4254b}).

\begin{figure}[ht]
                        \resizebox{\hsize}{!}{\includegraphics[angle=-90]{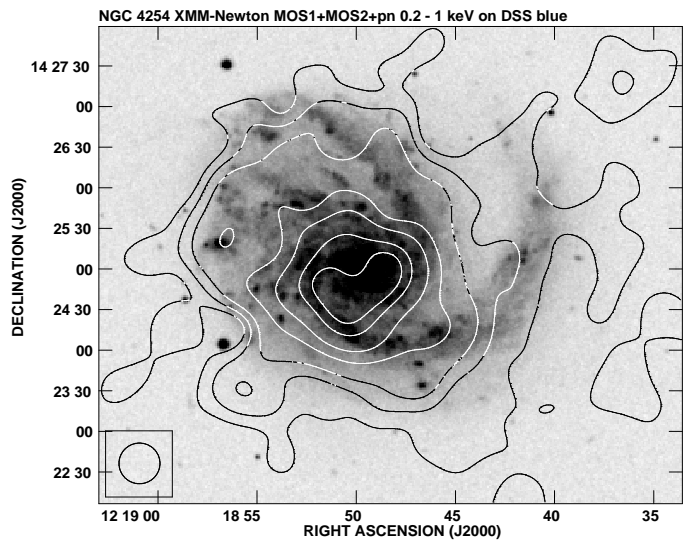}}
                \caption{
                XMM-Newton map of soft X-ray emission
                from NGC\,4254 in the 0.2 - 1 keV band overlaid on the Digital Sky Survey (DSS) blue image. The contours are 3, 5, 8, 16, 25, 40, and 50
                $\times$ rms. All point sources were subtracted. The map is convolved to the resolution of 30$\arcsec$. The smoothing scale is shown in the lower left corner of the figure.}
                \label{4254xfig}
        \end{figure}

The extended X-ray emission from NGC\,4254 is distributed uniformly (Fig.~\ref{4254xfig} and Chy\.zy et al.~\cite{chyzy4254}), with no enhancements present in the outer parts of the disk, especially
its southern area,
where the polarized ridge of the
radio emission is visible (Soida et al.~\cite{soida96}, Chy\.zy et al.~\cite{chyzy4254} and Fig.~\ref{4254radio}, see also Chy\.zy~\cite{chyzy4254b}).
%A bright peak visible in the southeastern
%part of the disk (Fig.~\ref{4254xfig}) is due to a strong point source, possibly an ultra luminous X-ray source. Such sources are often associated with colliding
%systems, as galactic interactions produce higher star formation rate, which may result in forming of more massive young
%X-ray binaries (see Soria \& Wong~\cite{soria}).
%This would be in agreement with the observations of NGC\,4254, suspected of tidal interactions (see Chy\.zy~\cite{chyzy4254b} and references therein).

\subsection{NGC\,4569}
\label{4569r}

%NGC\,4569 is an SABa galaxy situated relatively close to the Virgo cluster centre, at a distance of $1\fdg 66$ (500\,kpc in the sky plane)
%from Virgo A. It is an anaemic galaxy showing extended radio lobes unusual for
%a normal spiral (Chy\.zy et al.~\cite{chyzy4569}). At the southern egde of the western lobe, a ridge of polarized emission with the magnetic field parallel to it
%can be seen (see Fig.~\ref{4569radio}).
%For a detailed presentation and discussion of the radio data from this galaxy, see Chy\.zy et al.~(\cite{chyzy4569} and \cite{chyzy4569vla}).

\begin{figure}[ht]
\resizebox{\hsize}{!}{\includegraphics{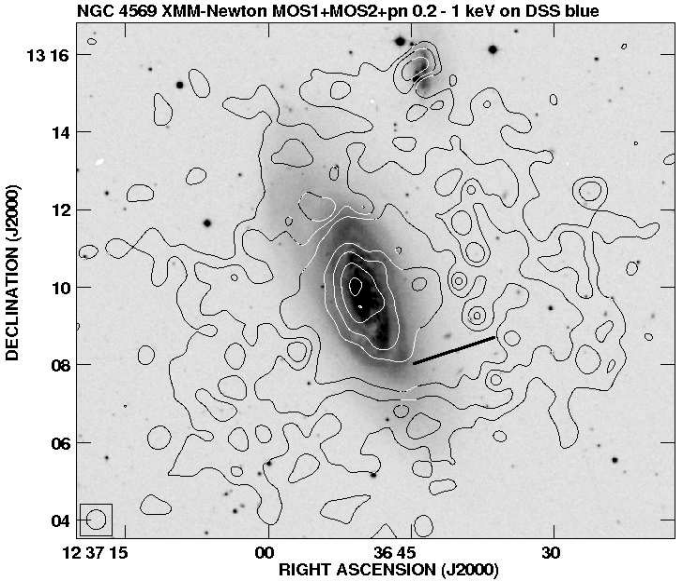}}
\caption{
XMM-Newton map of soft X-ray emission from NGC\,4569 in the 0.2 - 1 keV band overlaid on the DSS blue image. The contours are 3, 5, 8, 16, 25, 40, and 60
$\times$ rms. All point sources have been subtracted. The map has been convolved to the resolution of 30$\arcsec$. The smoothing scale is shown in the lower left corner of the figure. The solid line
marks the extension (see text).}
%{\it Right}: Regions of NGC\,4569 for which the spectra were aquired overlaid onto the DSS blue image. All background sources were extracted
%before creating each spectrum.}
\label{4569xfig}
\end{figure}

%The high resolution map of the X-ray emission (Fig.~\ref{4569xps})
%from this galaxy shows hot gas extensions visible on both sides of the optical disk. They reach up to 5\,kpc (1\arcmin) on the eastern and
%8.5\,kpc ($1\farcm 7$) on the western side and coincide with the observed radio features. 
For NGC\,4569, a low resolution map of the X-ray emission, which is more sensitive to extended structures, reveals
a large-scale hot gas halo around the galaxy, extending as far as 25\,kpc from the disk plane (Fig.~\ref{4569xfig}). North of the galaxy, still within the X-ray halo, lies an irregular galaxy,
IC\,3583, with which NGC\,4569 is suspected to interact (Tsch\"oke et al.~\cite{tschoke} and Chy\.zy et al.~\cite{chyzy4569}).
A slight X-ray extension (the third contour in Fig.~\ref{4569xfig} 
at R.A. = 12$^{\rm h}$36$^{\rm m}$38$^{\rm s}$, Dec = +13$\degr$ 08$\arcmin$) towards the region of the radio polarized ridge can be seen (marked in Fig.~\ref{4569xfig} with a solid line). 
Bright extensions close to the galactic disk coincide with the observed radio features, as
well as with H$\alpha$ outflows, which may provide more clues about the time evolution of this spatial coincidence after the decrease in the star forming activity (Bomans et al.~in prep.).

\subsection{NGC\,2276}
\label{2276obs}

%NGC\,2276 is an SABc galaxy in the NGC\,2300 group of galaxies. The optical structure is lopsided with a slight extension towards NGC\,2300 in the centre of the galaxy group. One of the spiral arms 
%seems to follow this extension. Star forming regions are somewhat shifted to the western and southwestern side of the galaxy, where a quite distinct and sharp edge is visible. 
%Radio observations by Hummel \& Beck~(\cite{humbeck}) show a steeper gradient of radio emission in this part of the galaxy. The authors notice here a ridge of polarized emission,
%with magnetic field parallel to the ridge,  
%as well as a slightly polarized radio tail extending to the east on the other side of the galaxy. 

\begin{figure}[ht]
                        \resizebox{\hsize}{!}{\includegraphics{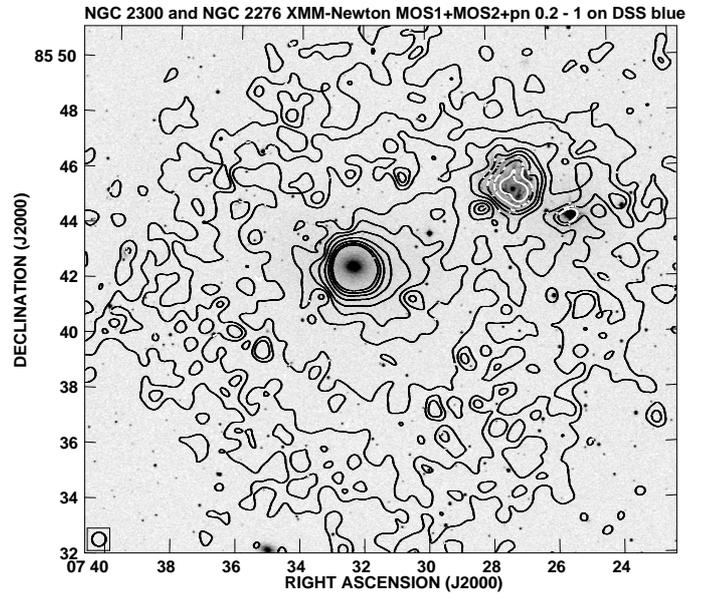}}
                \caption{XMM-Newton map of soft X-ray emission
                from hot gas cloud around NGC\,2276 and NGC\,2300 in the 0.2 - 1 keV band overlaid on the DSS blue image. The contours are 
		3, 5, 8, 13, 18, 25, 40, 60, 80, and 100 $\times$ rms. All point sources except for the central source of NGC\,2300 have been subtracted.
                The map has been convolved to the resolution of 30$\arcsec$. The smoothing scale is shown in the lower left corner of the figure.}
                \label{2300xfig}
        \end{figure}

In our wide-field map showing the hot gas halo around the NGC\,2300 group 
(Fig.~\ref{2300xfig}), an extension of hot gas from NGC\,2300 towards NGC\,2276 is clearly noticeable, on the same side of the disk of NGC\,2276 as the distorted southern spiral arm. 
NGC\,2276 shows a short tail of hot gas extending to the east. 

\begin{figure}[ht]
                        \resizebox{\hsize}{!}{\includegraphics{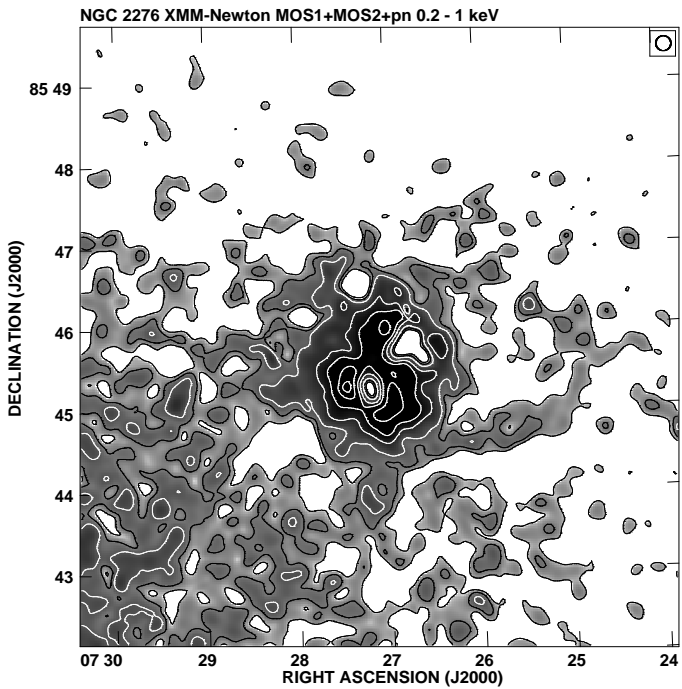}}
                \caption{
		XMM-Newton map of soft X-ray emission from NGC\,2276 in the 0.2 - 1 keV band (greyscale and contours). 
		The contours are 3, 5, 8, 16, 25, 40, and 50 $\times$ rms. All point sources have been subtracted. The map resolution is 10$\arcsec$. 
		The smoothing scale is shown in the upper right corner of the figure. 
			}	
                \label{2276xfig}
        \end{figure}

In contrast to the Chandra data analysed 
by Rasmussen et al.~(\cite{rasmus}), no strong gradient on the western side of the disk
is visible (Fig.~\ref{2276xfig}). The higher resolution image of NGC\,2276 (Fig.~\ref{2276xfig}) shows, in contrast to the corresponding map in Rasmussen et al.~(\cite{rasmus}),
numerous 3$\sigma$ patches of emission north and west of the galaxy, which is clear evidence of soft X-ray emission of low surface-brightness, that might not have been detected 
by Chandra, owing to its lower sensitivity. This emission is visible in the wide-field low resolution map in Fig.~\ref{2300xfig}. This could therefore also 
lead to the ''flattening'' of the emission gradient observed on the western edge of the disk.

%This does not seem to be a resolution effect, as the emission gradient seen in the Chandra image (Rasmussen et al.~\cite{rasmus}) spans over some 0$\farcm$5 and our 
%image has a resolution of 10'' (Fig.~\ref{2276xfig} left). Therefore, if a strong gradient existed in our data, we could still detect it. 
%One can however see a somewhat asymmetrical distribution of the hot gas outside of the disk of NGC\,2276 in the low resolution image (Fig.~\ref{2276xfig} right). 
%Provided the galaxy is moving to the west, this could be an argument for gas sweeping by the galaxy.  

\section{Spectral analysis of the X-ray emission}
\label{spectra}

We now examine the properties of the hot gas in the studied galaxies. Spectra of 
the X-ray emission were obtained for selected regions of the galactic disk in the case of NGC\,4254, and for regions corresponding to H$\alpha$ outflows and radio lobes in the case of NGC\,4569.
For the XMM-Newton observations of NGC\,2276, we repeated part of the analysis performed by Rasmussen et al.~(\cite{rasmus}), 
who examined the possible bow shock at the western edge of the disk and the corresponding hot gas tail extending to the east, 
as well as determined the temperature of the IGM around the galaxy. The only change we introduce here is a separate analysis of the region, where a radio polarized ridge is visible
(Hummel \& Beck~\cite{humbeck}). This allows us to obtain consistent data that can be easily compared with the two other cases (NGC\,4254 and NGC\,4569).

For all spectra investigated by us, we used a model composed of a thermal plasma and a contribution from unresolved point-like sources, 
as well as possible residuals from the excised point sources, when necessary.
Thermal plasma is represented in this work by a {\it mekal} model, which is an emission spectrum from hot diffuse gas based on the model calculations of Mewe and Kaastra~(Mewe et al.~\cite{mewe},
Kaastra~\cite{kaastra}). The metallicity is always fixed to solar, except for the analysis of the data for NGC\,2276 (see Sect.~\ref{2276spec}). 
A contribution from unresolved point-like sources is fitted with a simple power-law. Owing to the limited resolution of the XMM-Newton EPIC cameras, this component is needed for most of the presented 
spectra, both within a galactic disk (X-ray binaries) and in its surroundings (background sources). 
In our models, we use single- or two-temperature thermal plasmas, depending on the expected
physical conditions of the emission and/or on the single-peak or double-peak appearance of the spectrum. A two-temperature model seems to be a good approach in the case of studying 
outflows of the hot gas from the galactic disk (e.g. T\"ullmann et al.~\cite{tullmann06}), as it is the simplest approximation of the expected multi-temperature plasma, 
when the outflow gas mixes into the galactic halo. Adding more plasma components would require much more
sensitive data. The errors provided for all model parameters are always 1$\sigma$ errors.

Since we analyse the emission from cluster/group galaxies, we can expect a contribution of the ICM/IGM to the source spectra, which ideally should be subtracted. However, owing to 
low surface-brightness of this emission, this contribution is expected to be very low within the small regions used in our analysis. The moderate sensitivity 
of the data used requires large areas to obtain usable spectra of the ICM/IGM emission (as in the case of NGC\,2276). Furthermore, since the data that we used do not allow us to 
obtain precisely constrained parameters of the ICM/IGM, subtraction of the emission for this component from the source spectra would most likely cause large uncertainties, as a 
consequence of the error propagation. 

Another difficulty might arise from the small extent of some spectral regions (e.g. regions 2 and 3 for NGC\,2276), as this may lead to a point spread function (PSF) 
leakage of photons between adjacent areas. 
However, we consider this effect to be negligible, as in our regions we do not expect strong gradients on length scales comparable to the PSF, with the studied regions being still several times larger.  

As already mentioned in Sect.~\ref{obsred}, the datasets for NGC\,4254 and NGC\,4569 contain soft proton contamination. To keep the present sensitivity of our data, instead of 
applying further filtering we checked the possible influence of this contamination using the most sensitive spectra for both datasets. As recommended by Kuntz \& Snowden~(\cite{kuntz08}), we fitted 
these spectra as presented in Sects.~\ref{4254spec} and \ref{4569spec}, but added a broken power-law component to account for the soft proton contamination. We compared the resulting model 
fits with the original ones, which showed that the temperature values did not change and that the corresponding fluxes still agreed within the errors with those derived from model fits without the 
broken power-law component. We therefore concluded, that the soft proton contamination of the datasets of NGC\,4254 and NGC\,4569 is so low, that the fit parameters for thermal plasma 
derived from the spectral analysis presented below do not change. Although an obvious change is observed for the power-law component, this remains of minor importance for this paper. 

\subsection{NGC\,4254}
\label{4254spec}

\begin{figure}[ht]
\resizebox{\hsize}{!}{\includegraphics{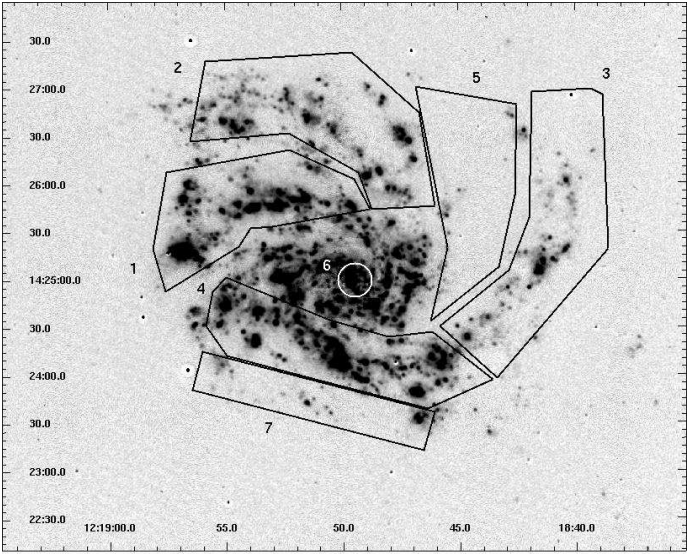}}
\caption{Regions of NGC\,4254 for which the spectra were acquired overlaid on the H$\alpha$ image.
All background sources were excised before extracting each spectrum.}
\label{4254xreg}
\end{figure}

\begin{figure*}[ht]
                \resizebox{5cm}{!}{\includegraphics[angle=-90]{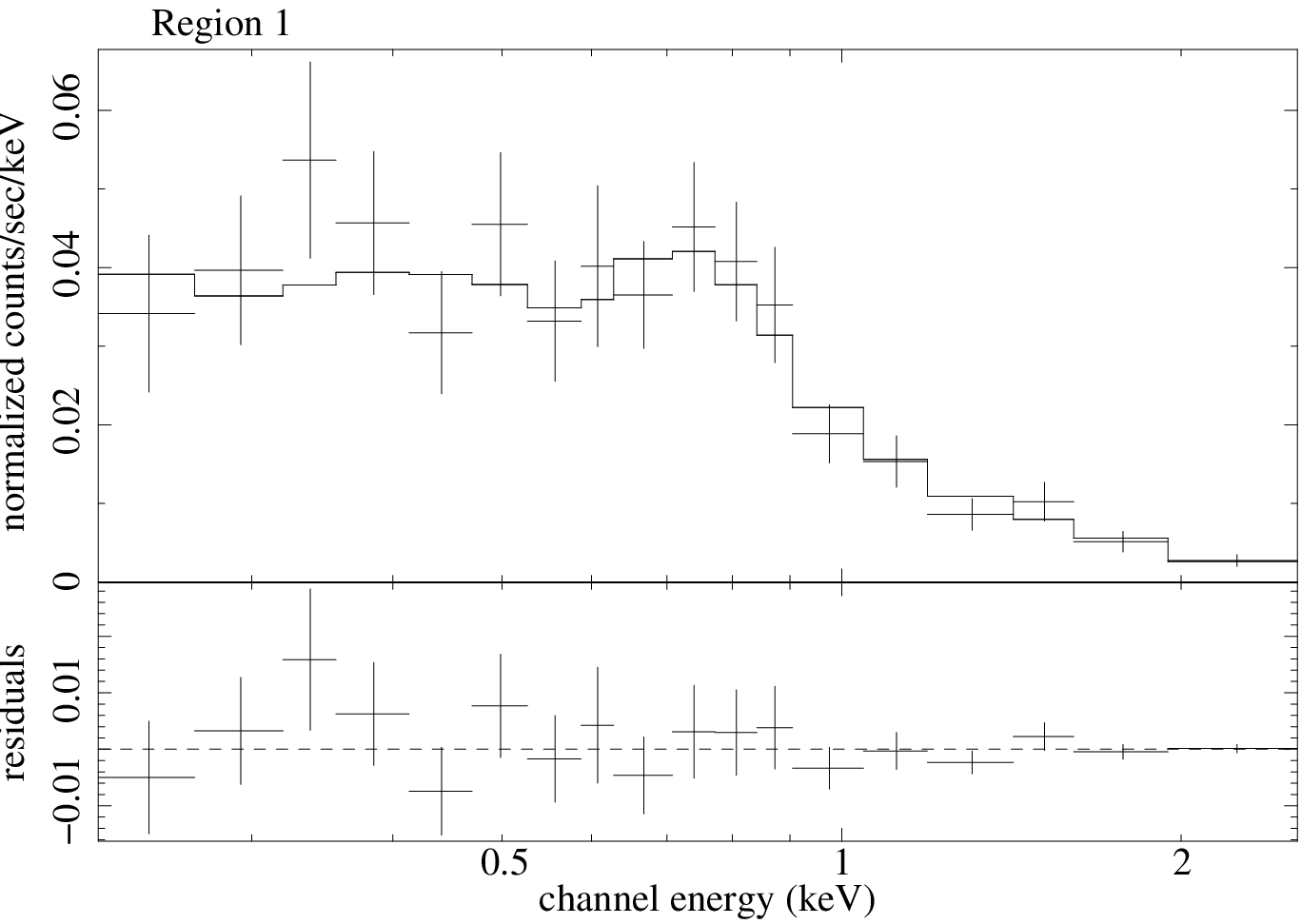}}
                \resizebox{5cm}{!}{\includegraphics[angle=-90]{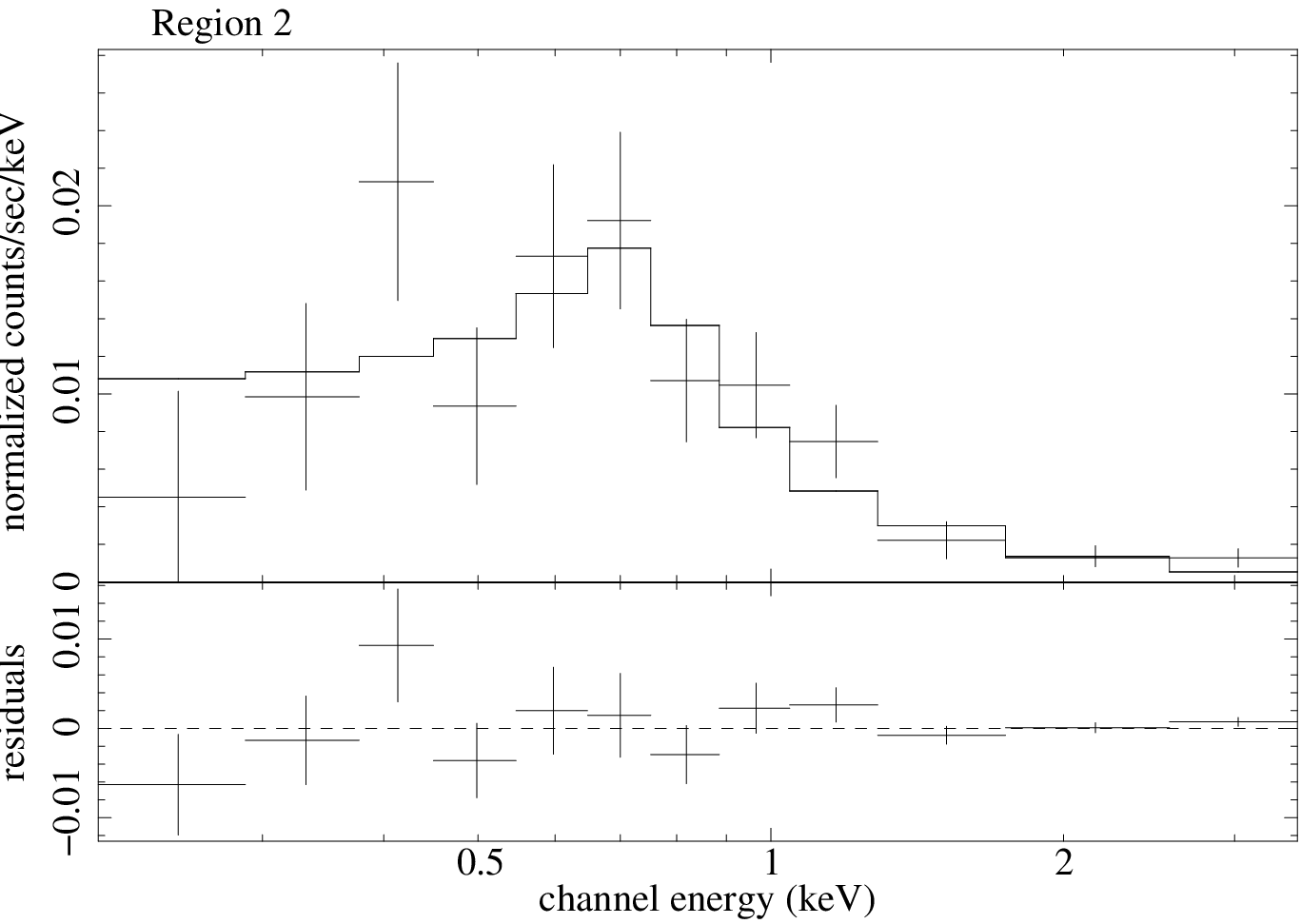}}
                \resizebox{5cm}{!}{\includegraphics[angle=-90]{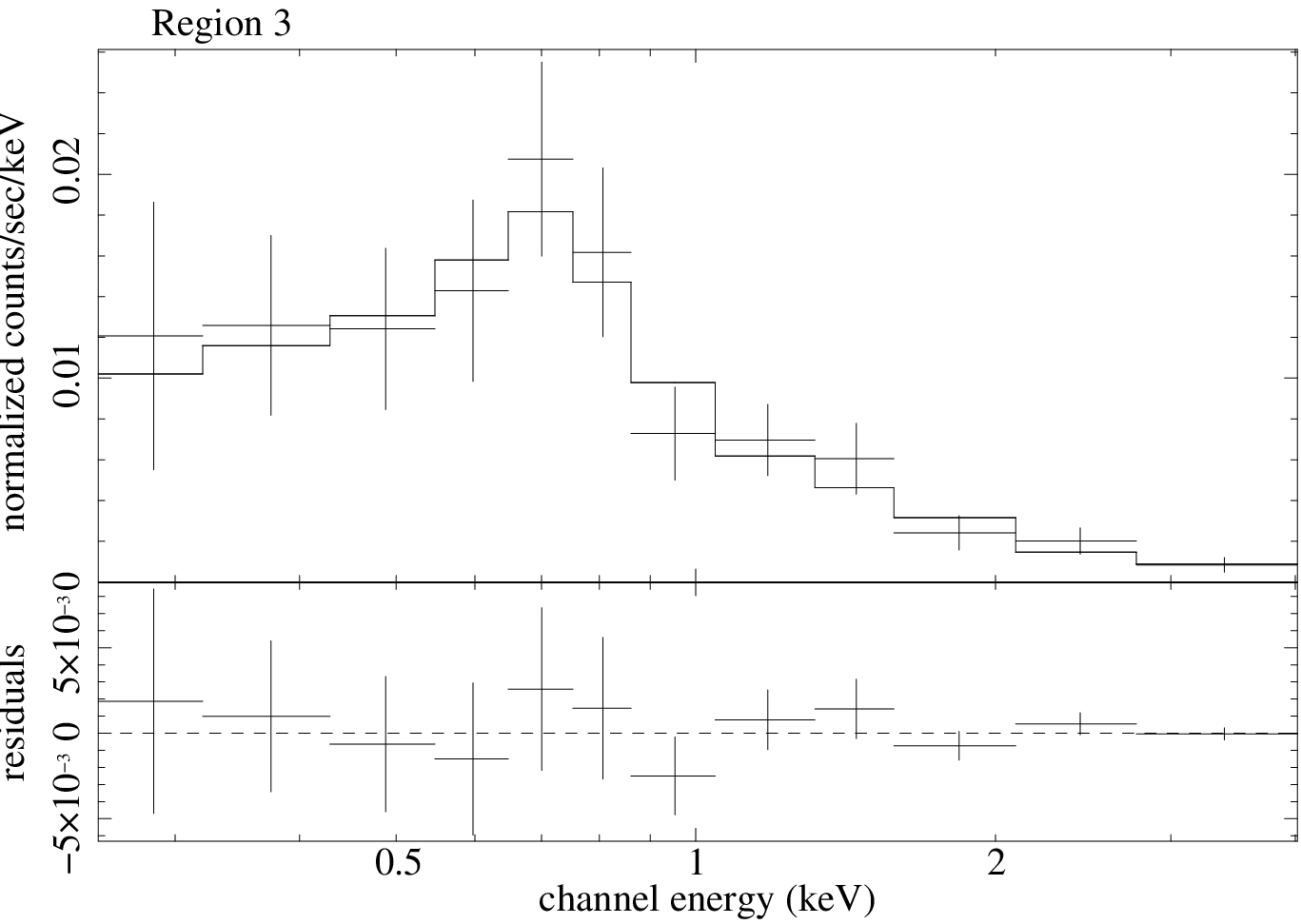}}
                \resizebox{5cm}{!}{\includegraphics[angle=-90]{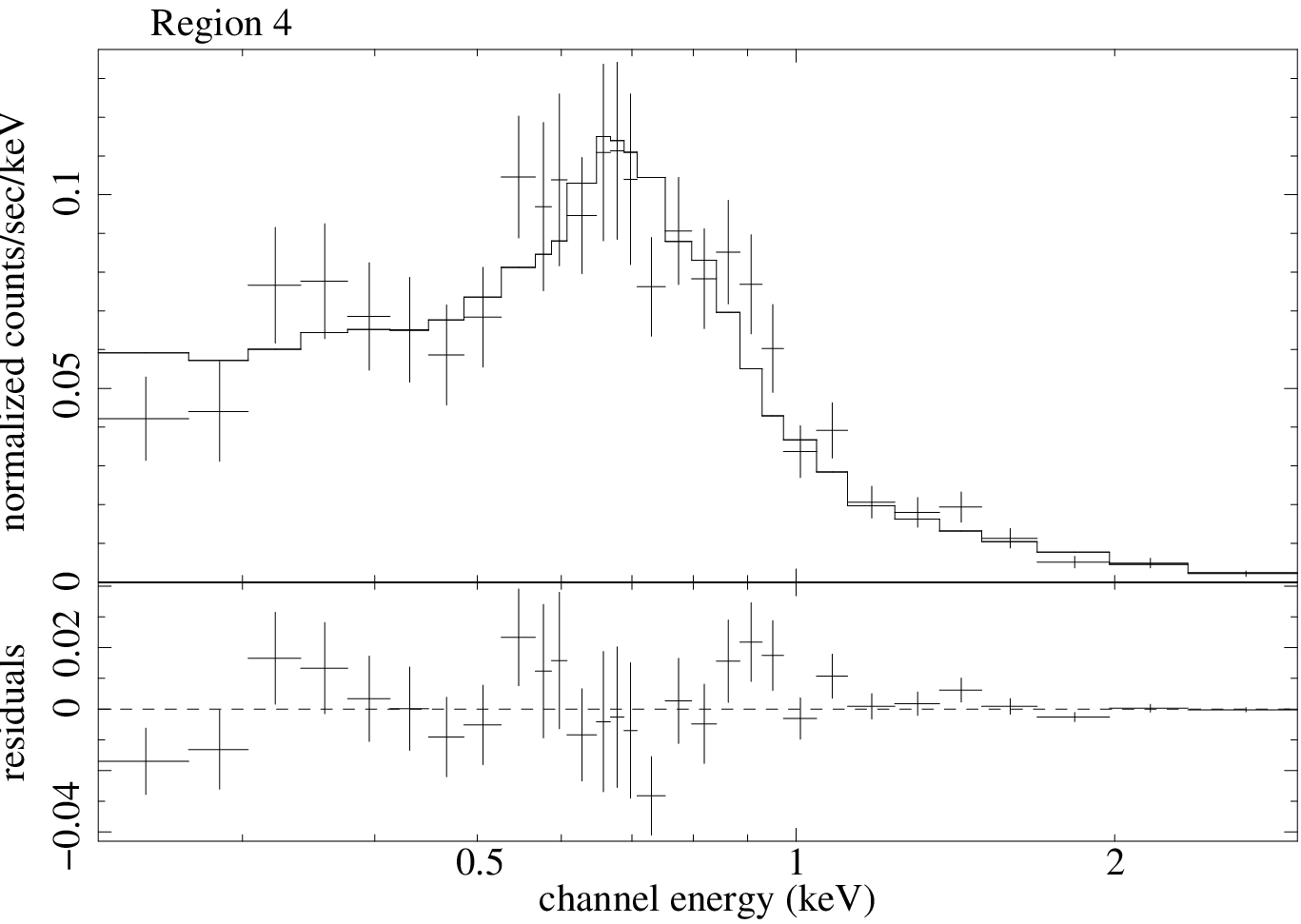}}
                \resizebox{5cm}{!}{\includegraphics[angle=-90]{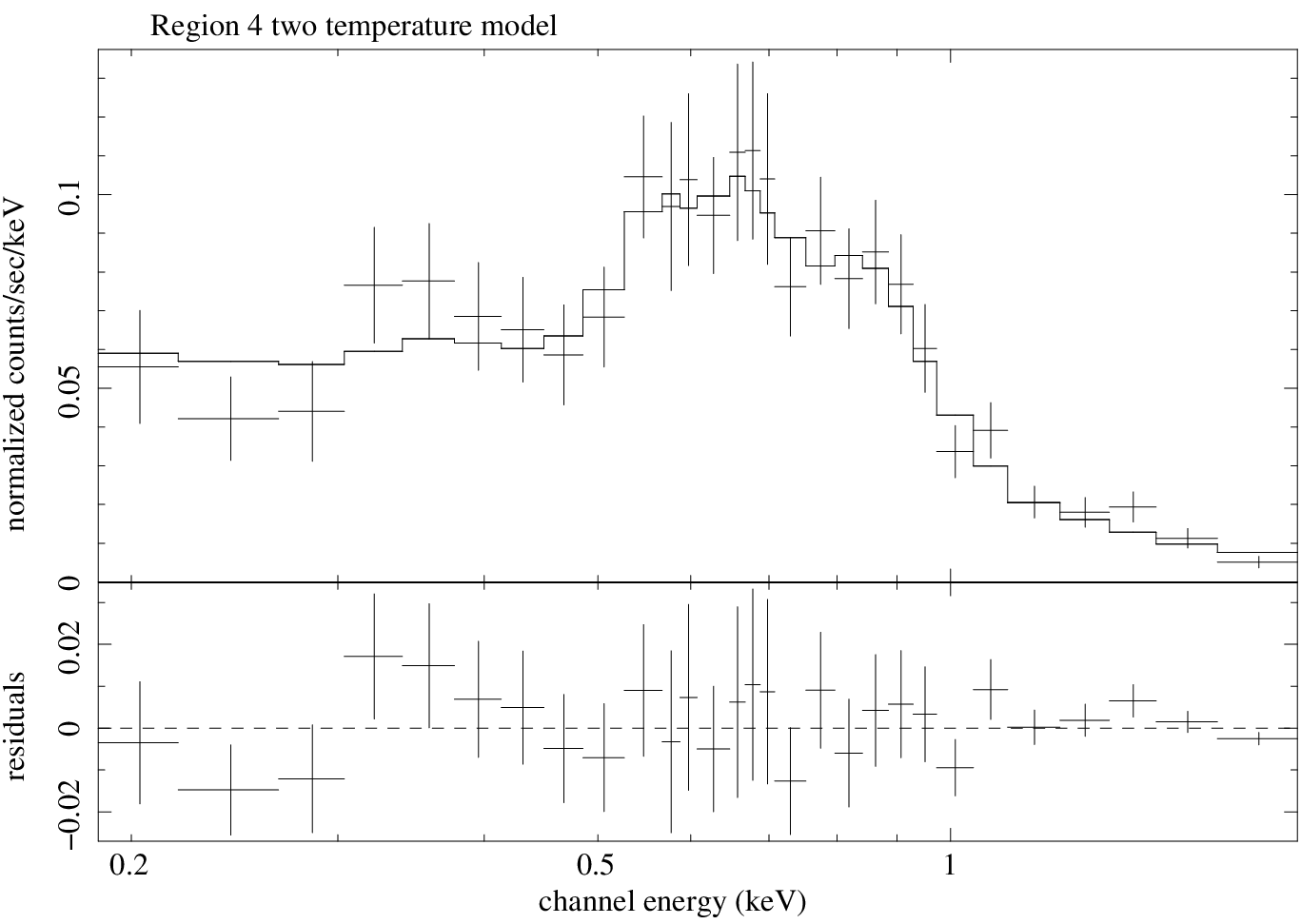}}
                \resizebox{5cm}{!}{\includegraphics[angle=-90]{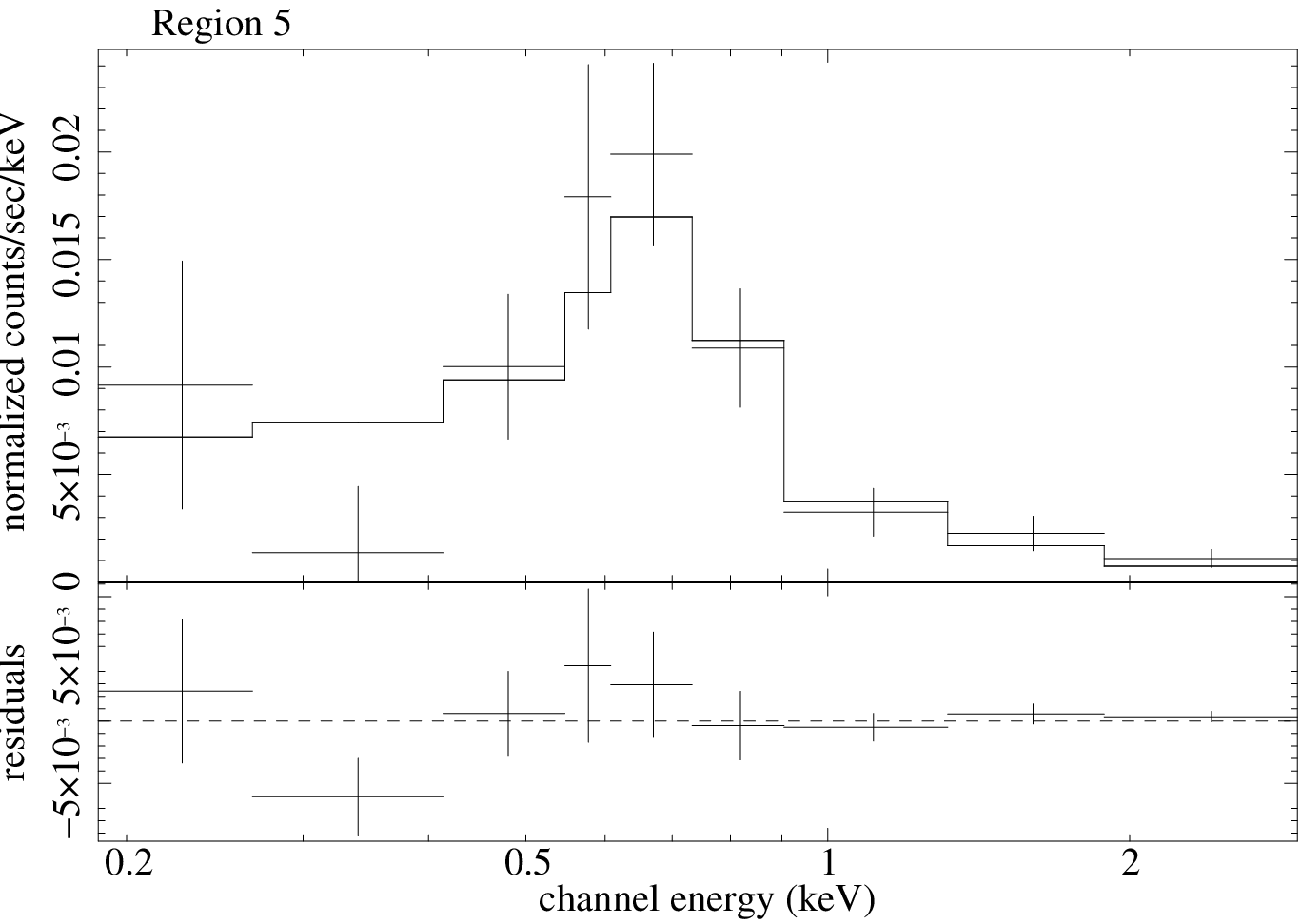}}
                \resizebox{5cm}{!}{\includegraphics[angle=-90]{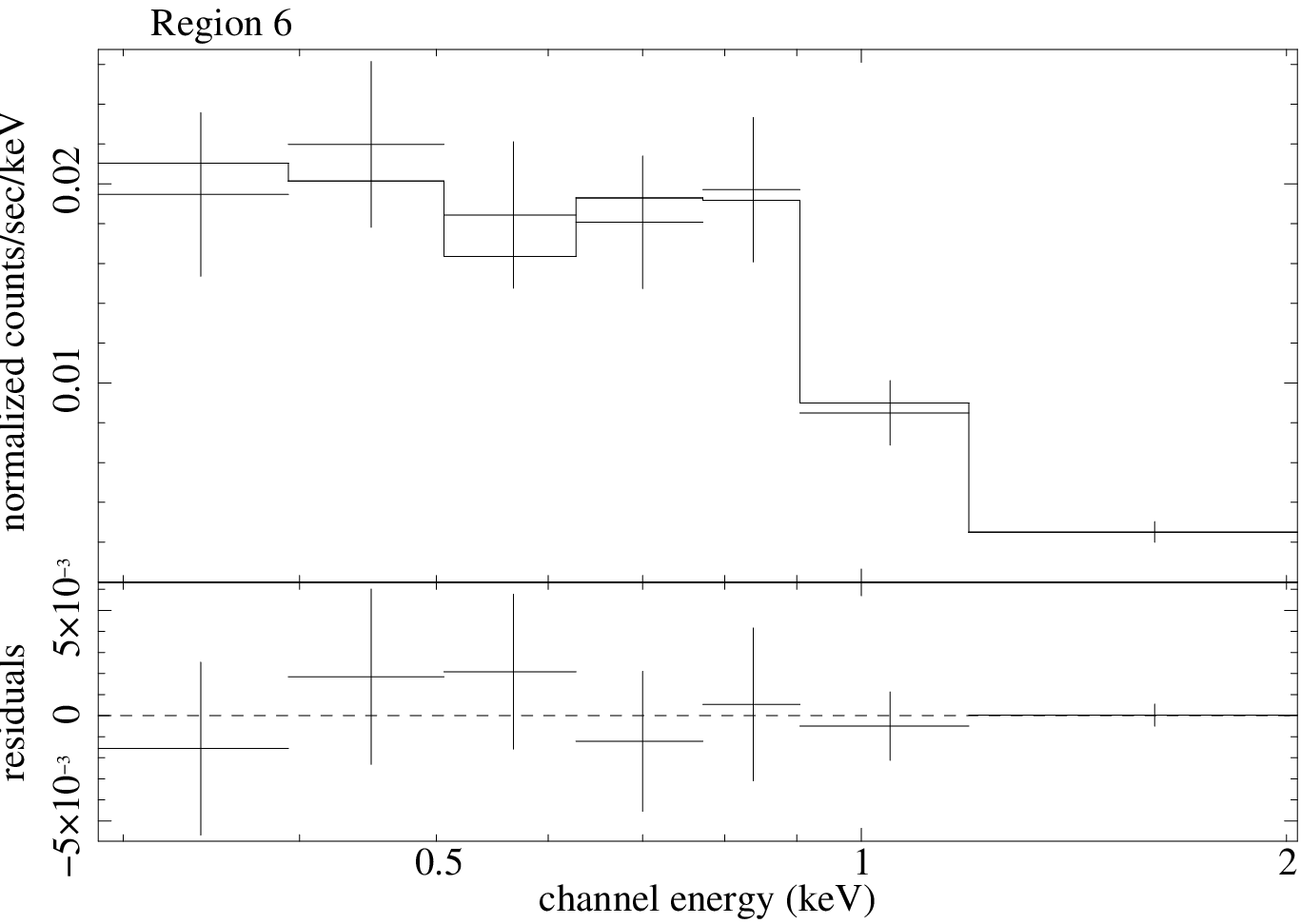}}
                \resizebox{5cm}{!}{\includegraphics[angle=-90]{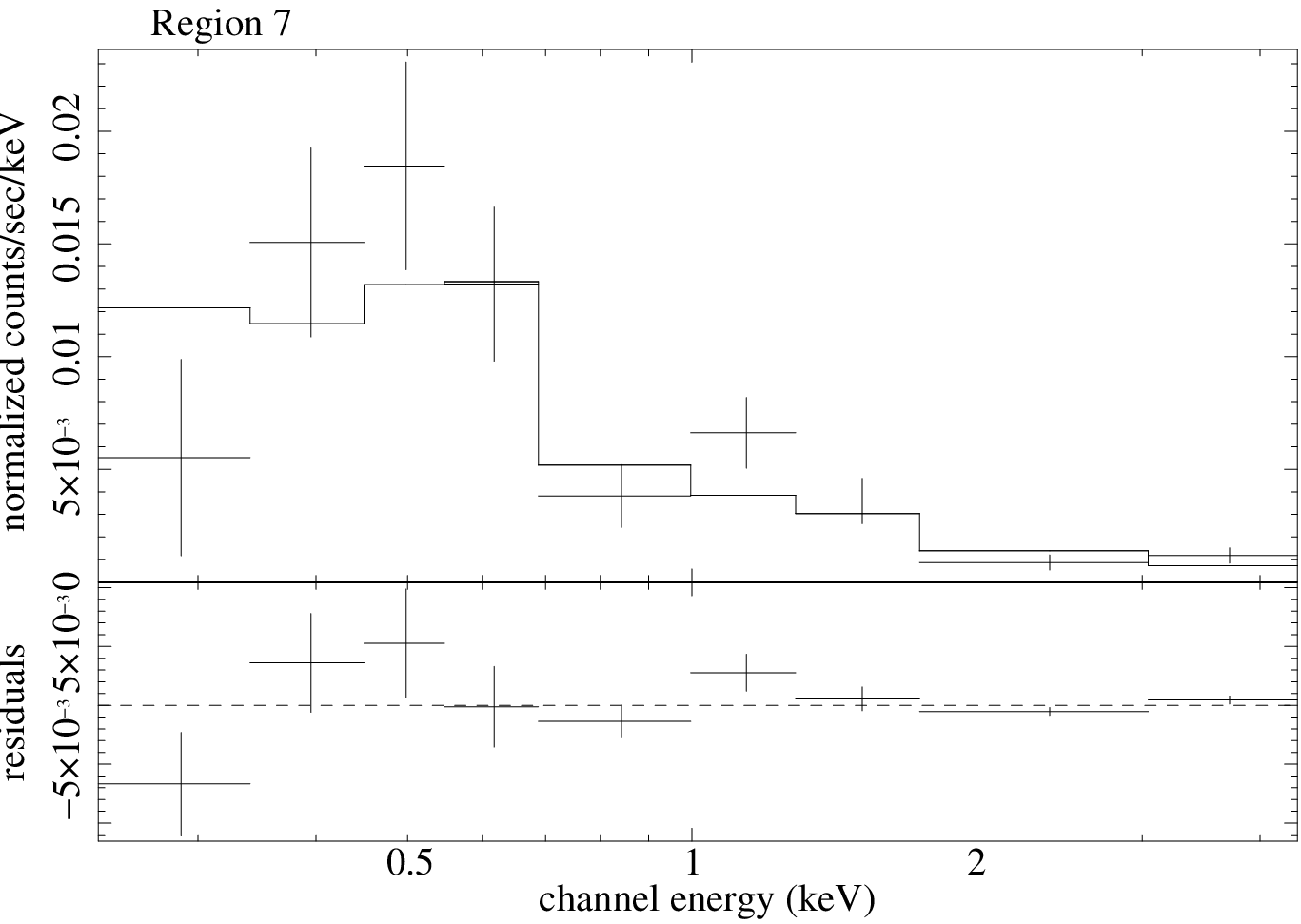}}
        \caption{Model fits to the regions of NGC\,4254. See Tables~\ref{4254xtab} and
        \ref{4254xf}.}
        \label{4254mod}
\end{figure*}

The relatively uniform hot gas distribution in the disk of NGC\,4254 does not suggest that there are any compressions or outflows, however 
the spectral analysis reveals more clues. All regions presented in Fig.~\ref{4254xreg} were
fitted with a single-temperature thermal plasma with
an admixture of a power-law to account for unresolved point sources. 
%In case of the polarized ridge region (region 4 in Fig.~\ref{4254xreg}) 
%a better statistics than in the other spectra allowed us also
%a two temperature thermal plasma model fit.
Galactic spiral arms (regions 1, 2, and 3) showed similar gas temperatures of the order of 0.3~keV (see Table~\ref{4254xtab} for details).
Hot gas in the region of a radio polarized ridge (region 4) was found to have the same temperature as the gas in the spiral arms, again around 0.3~keV. 
We also checked all regions for a hotter gas component by fitting a two-temperature plasma model. 
Such a model can be explained as emission from a mixture of two thermal plasmas before reaching a thermal equilibrium. We found hints of a second component 
only for region 4, where the polarized radio ridge is visible.
The two-temperature model fit resulted in
temperatures of 0.22$^{+0.05}_{-0.03}$~keV and 0.64$^{+0.12}_{-0.11}$~keV. We need to note, however, that this is a spectrum extracted from the region with the highest quality photon statistics. 
A hotter component in the emission from NGC\,4254 might exist, but is difficult to trace in the spectra for other regions of this galaxy due to lower quality of the statistics. 
At the edge of the polarized ridge (region 7), the coolest gas was found, with a temperature of 0.14$\pm$0.04~keV. In the nuclear region (region 6), the temperature 
of the hot gas reaches 0.41$^{+0.29}_{-0.16}$~keV. The spectra for all regions together with best-fit models are presented in Fig.~\ref{4254mod}. 
The fit parameters and derived fluxes are presented in Tables \ref{4254xtab} and \ref{4254xf}. A thorough discussion of the fits is presented in Sect.~\ref{4254dis}.

\begin{table}[ht]
\caption{\label{4254xtab}Model fit parameters of NGC\,4254 regions.}
\centering
\begin{tabular}{llllll}
\hline\hline
Reg.      & kT$_1$    &  kT$_2$    & Photon    & Net &  Red.  \\%& nhp$^{\ast}$ \\
\vspace{5pt} no. & [keV]  &  [keV]  &  Index  &  counts\tablefootmark{a} & $\chi^2$  \\%&  \\
\hline
\vspace{5pt}
1 & 0.37$^{+0.16}_{-0.10}$ & --                         & 2.13$^{+0.20}_{-0.22}$  & 484  & 0.5     \\%& 0.954 \\
\vspace{5pt}
2 & 0.30$^{+0.24}_{-0.07}$ & --                         & 1.75$^{+0.43}_{-0.46}$  & 184  & 1.2     \\%& 0.252 \\
\vspace{5pt}
3 & 0.29$^{+0.21}_{-0.08}$ & --                         & 1.43$^{+0.49}_{-0.59}$  & 227  & 0.5     \\%& 0.830 \\
\vspace{5pt}
4 & 0.29$\pm$0.03          & --                         & 1.88$^{+0.21}_{-0.23}$  & 953  & 1.2     \\%& 0.262 \\
\vspace{5pt}
  & 0.22$^{+0.05}_{-0.03}$ & 0.64$^{+0.12}_{-0.11}$     & 1.69$^{+0.28}_{-0.31}$  & 953  & 0.8       \\%& 0.786 \\
\vspace{5pt}
5 & 0.27$^{+0.07}_{-0.04}$ & --                         & 1.11$^{+0.51}_{-0.38}$  & 135  & 0.9     \\%& 0.536 \\
\vspace{5pt}
6 & 0.41$^{+0.29}_{-0.16}$ & --                         & 2.34$^{+0.32}_{-0.37}$  & 189  & 0.9       \\%& 0.464 \\
\vspace{5pt}
7 & 0.14$^{+0.04}_{-0.03}$ & --                         & 1.16$^{+1.41}_{-0.35}$  & 165  & 1.9     \\%& 0.069 \\
\hline
\end{tabular}
\tablefoot{
\tablefoottext{a}{Total net counts in the 0.2-12 keV band.}
}
\end{table}

\begin{table}[ht]
        \caption{\label{4254xf}Total (0.2 - 12 keV) and soft (0.2 - 1 keV) unabsorbed fluxes in 10$^{-14}$erg\,cm$^{-2}$s$^{-1}$ for the modelled regions in NGC\,4254.}
\centering
\begin{tabular}{lllll}
\vspace{5pt} Reg. no.  & mekal 1 & mekal 2 & powerlaw  & total \\
\hline\hline
\vspace{5pt}
1 (total) & 1.5$^{+1.2}_{-0.9}$ & -- & 11$^{+3}_{-2}$ & 13$^{+4}_{-3}$ \\
1 (soft)  & 1.3$^{+0.9}_{-0.7}$ & -- & 5.2$^{+1.9}_{-1.5}$ & 6.5$\pm$0.7 \\
\vspace{5pt}
2 (total) & 1.0$^{+0.9}_{-0.7}$ & -- & 4$^{+6}_{-2}$ & 5$^{+7}_{-3}$ \\
2 (soft) & 0.9$^{+0.6}_{-0.7}$ & -- & 1.2$\pm$0.1 & 2.1$^{+0.6}_{-0.7}$ \\
\vspace{5pt}
3 (total) & 1.0$^{+1.1}_{-0.8}$ & -- & 8$^{+14}_{-5}$ & 9$^{+15}_{-6}$ \\
3 (soft) & 0.9$\pm$0.8 & -- & 1.3$^{+1.3}_{-0.7}$ & 2.2$\pm$0.8 \\
\vspace{5pt}
4 (total) & 5.7$\pm$1.2	        & -- & 13$^{+5}_{-2}$ & 19$^{+6}_{-3}$ \\
4 (soft) & 5.2$^{+0.9}_{-1.1}$ & -- & 4.2$^{+1.7}_{-1.1}$ & 9.4$^{+0.9}_{-1.1}$ \\
\vspace{5pt}
(total)  & 4.5$^{+1.7}_{-1.5}$ & 3$^{+1}_{-2}$ & 12$^{+9}_{-4}$ & 20$^{+12}_{-8}$ \\
(soft) & 4.4$^{+1.4}_{-1.5}$ & 2.2$^{+1}_{-1.3}$ & 2.9$^{+1.7}_{-1}$ & 9.5$^{+4.1}_{-3.8}$\\  
\vspace{5pt}
5 (total) & 1.2$^{+0.5}_{-0.4}$ & -- & 4$^{+8}_{-3}$ & 5$^{+9}_{-3}$ \\
5 (soft) & 1.1$\pm$0.4 & -- & 0.3$^{+0.5}_{-0.2}$ & 1.4$^{+0.8}_{-0.6}$\\  
\vspace{5pt}
6 (total) & 0.6$^{+0.6}_{-0.4}$ & -- & 4$\pm$1 	  & 5$^{+2}_{-1}$ \\
6 (soft) & 0.5$^{+0.5}_{-0.3}$ & -- & 2.2$^{+1.3}_{-1}$ & 2.7$^{+1.8}_{-1.3}$\\  
\vspace{5pt}
7 (total) & 1.0$^{+1.2}_{-0.7}$ & -- & 6$^{+8}_{-4}$ & 7$^{+9}_{-4}$ \\
7 (soft) & 1.0$^{+1.2}_{-0.7}$ & -- & 0.5$^{+0.4}_{-0.3}$ & 1.5$^{+1.6}_{-0.9}$\\ 
\hline
\end{tabular}
\end{table}

\subsection{NGC\,4569}
\label{4569spec}

\begin{figure}[ht]
\resizebox{\hsize}{!}{\includegraphics{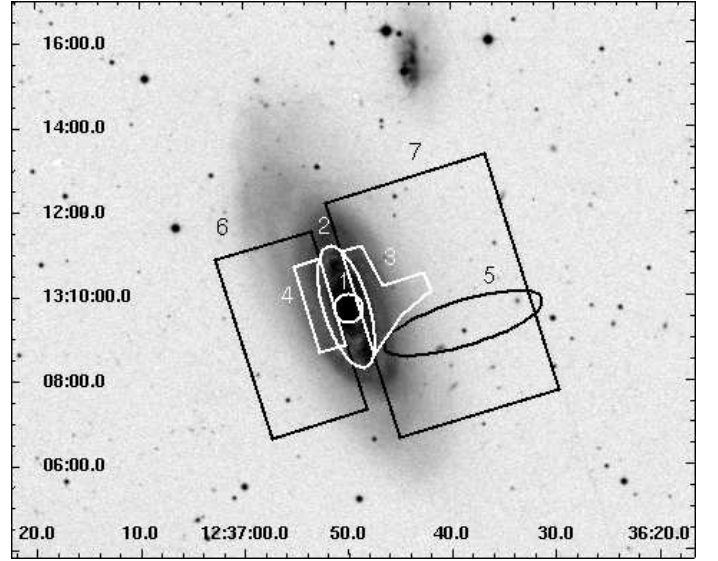}}
\caption{Regions of NGC\,4569 for which the spectra were aquired overlaid on the DSS blue image. Each area is used for the spectral analysis only once (e.g. region 4 is not a part of
region 6). All background sources were excised before extracting each spectrum.}
\label{4569xreg}
\end{figure}

\begin{figure*}[ht]
\resizebox{5cm}{!}{\includegraphics[angle=-90]{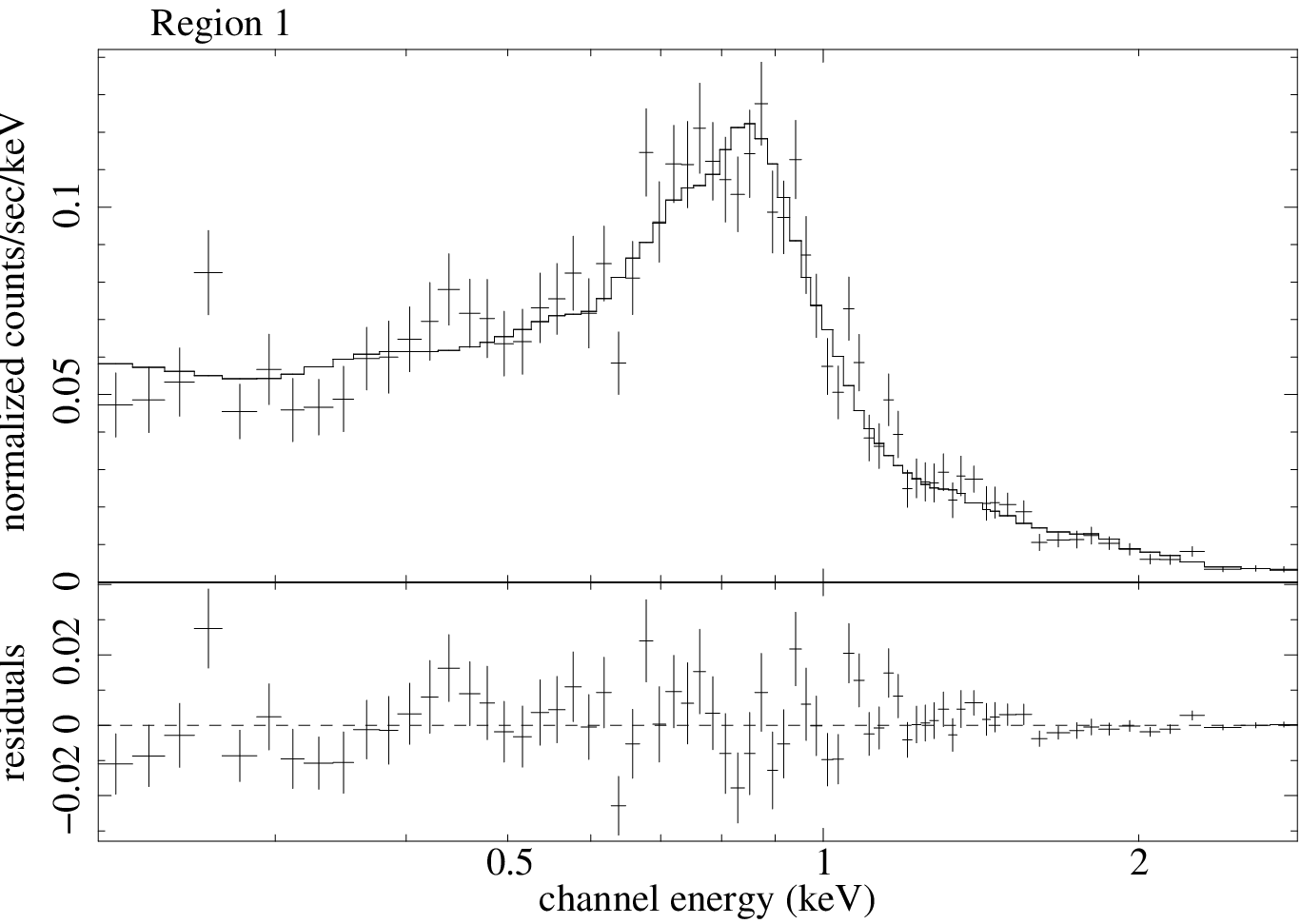}}
\resizebox{5cm}{!}{\includegraphics[angle=-90]{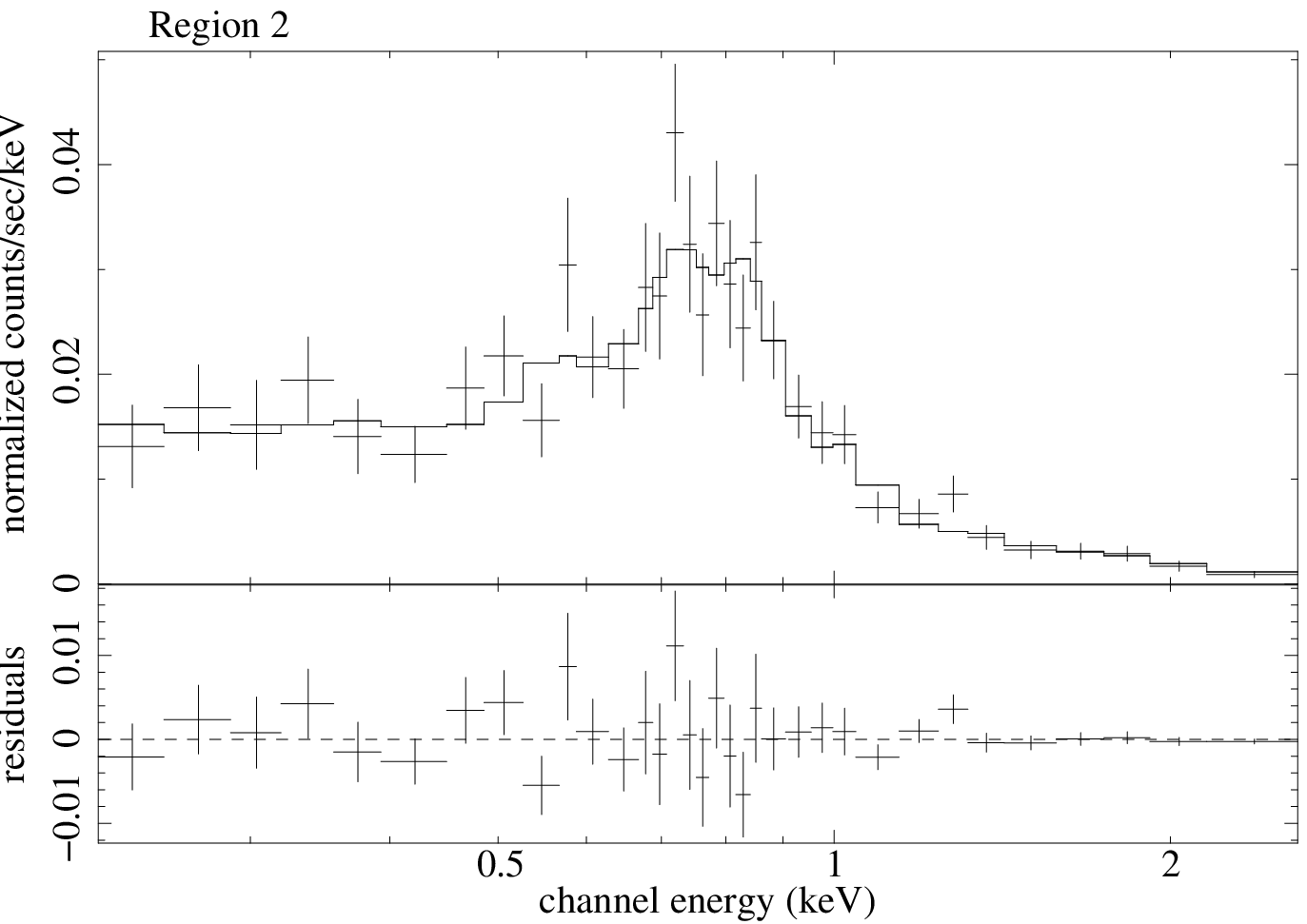}}
\resizebox{5cm}{!}{\includegraphics[angle=-90]{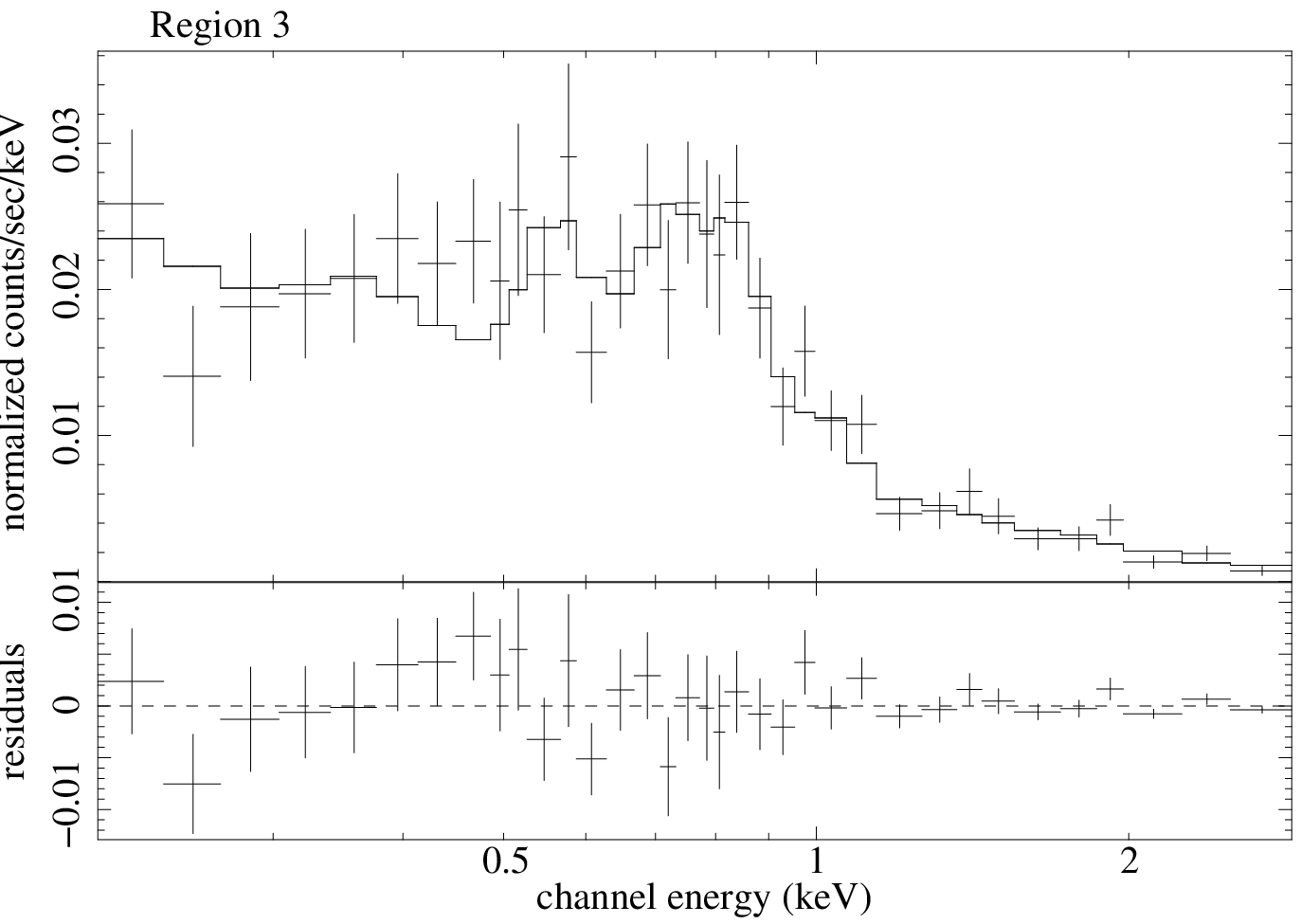}}
\resizebox{5cm}{!}{\includegraphics[angle=-90]{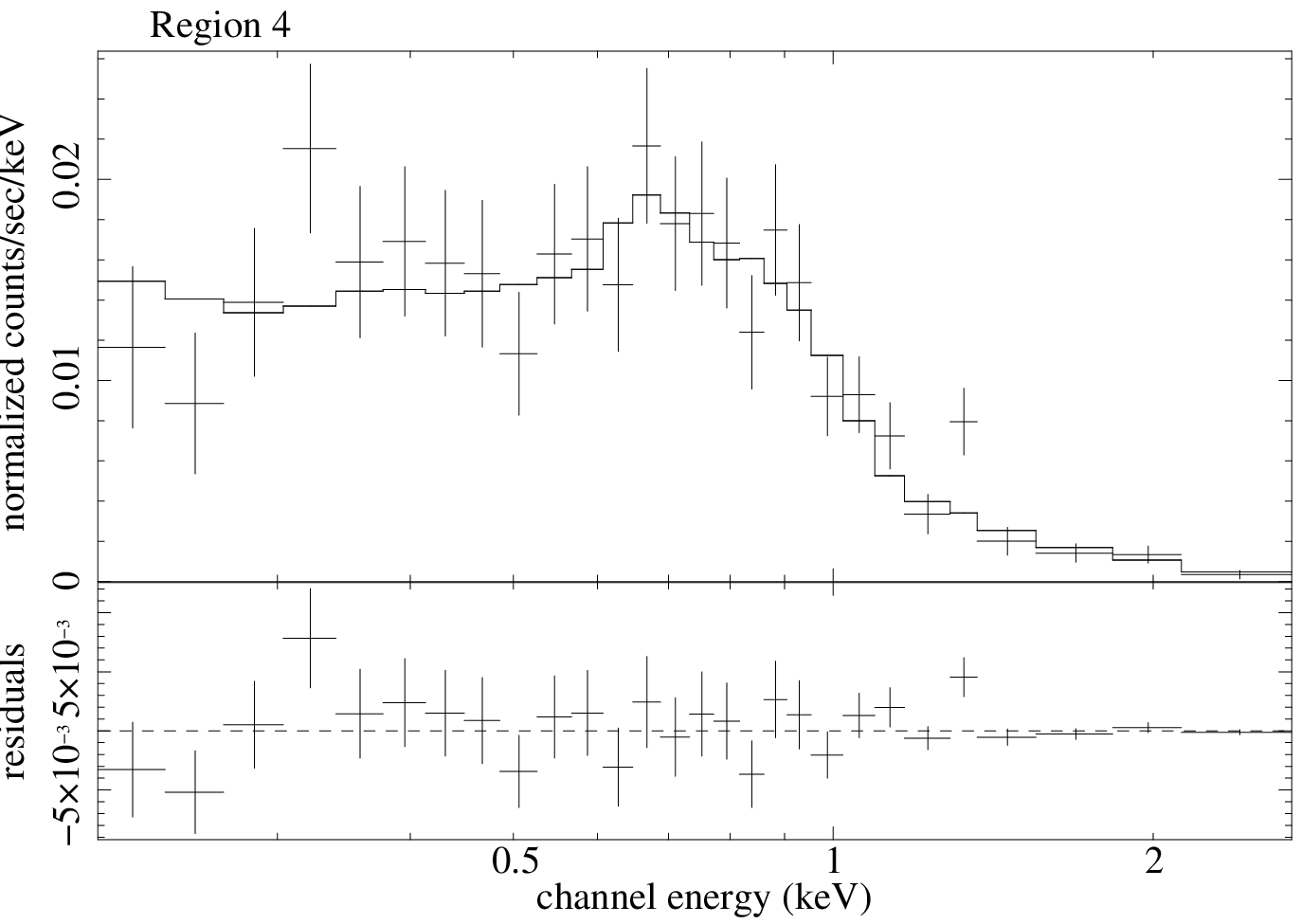}}
\resizebox{5cm}{!}{\includegraphics[angle=-90]{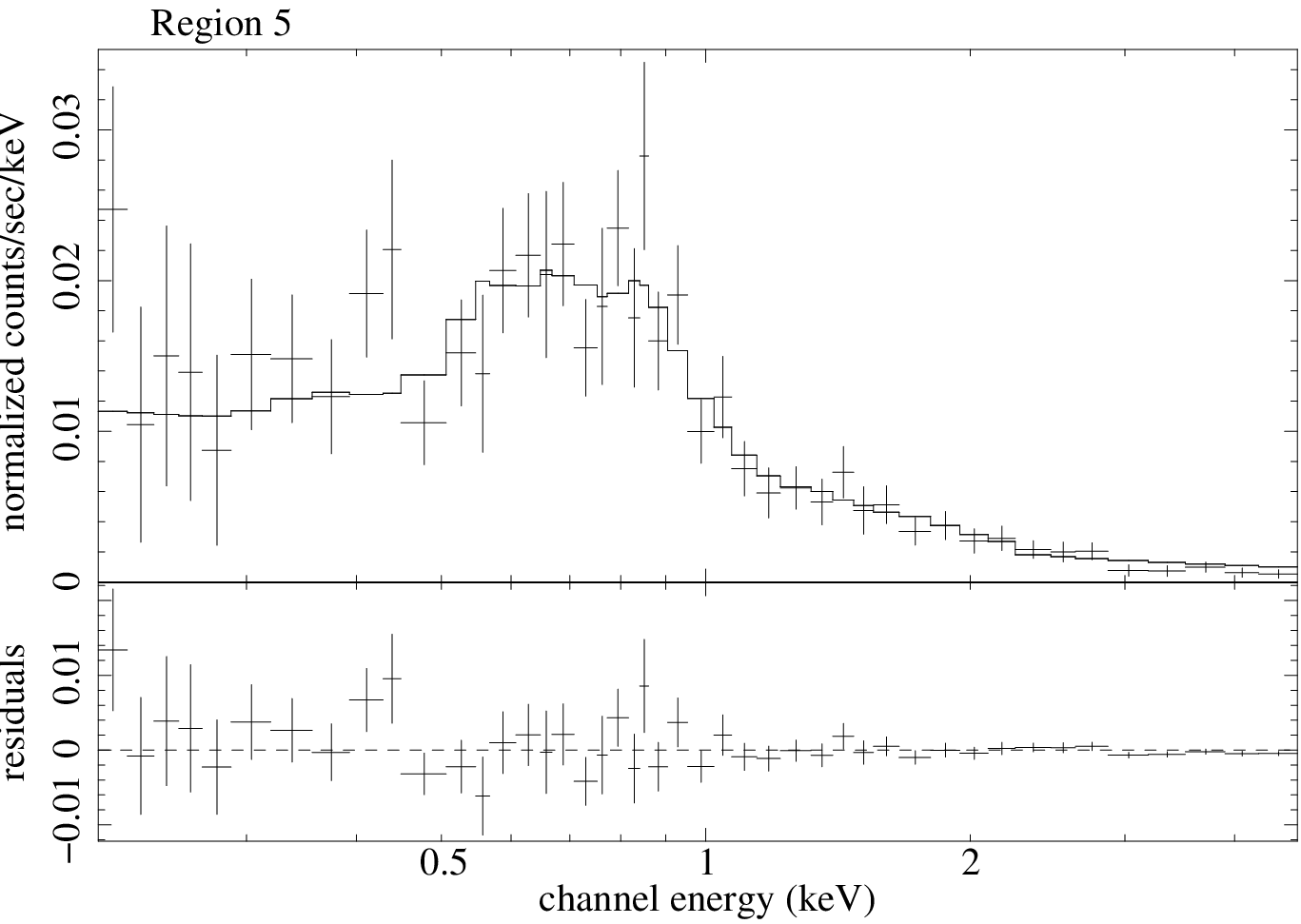}}
\resizebox{5cm}{!}{\includegraphics[angle=-90]{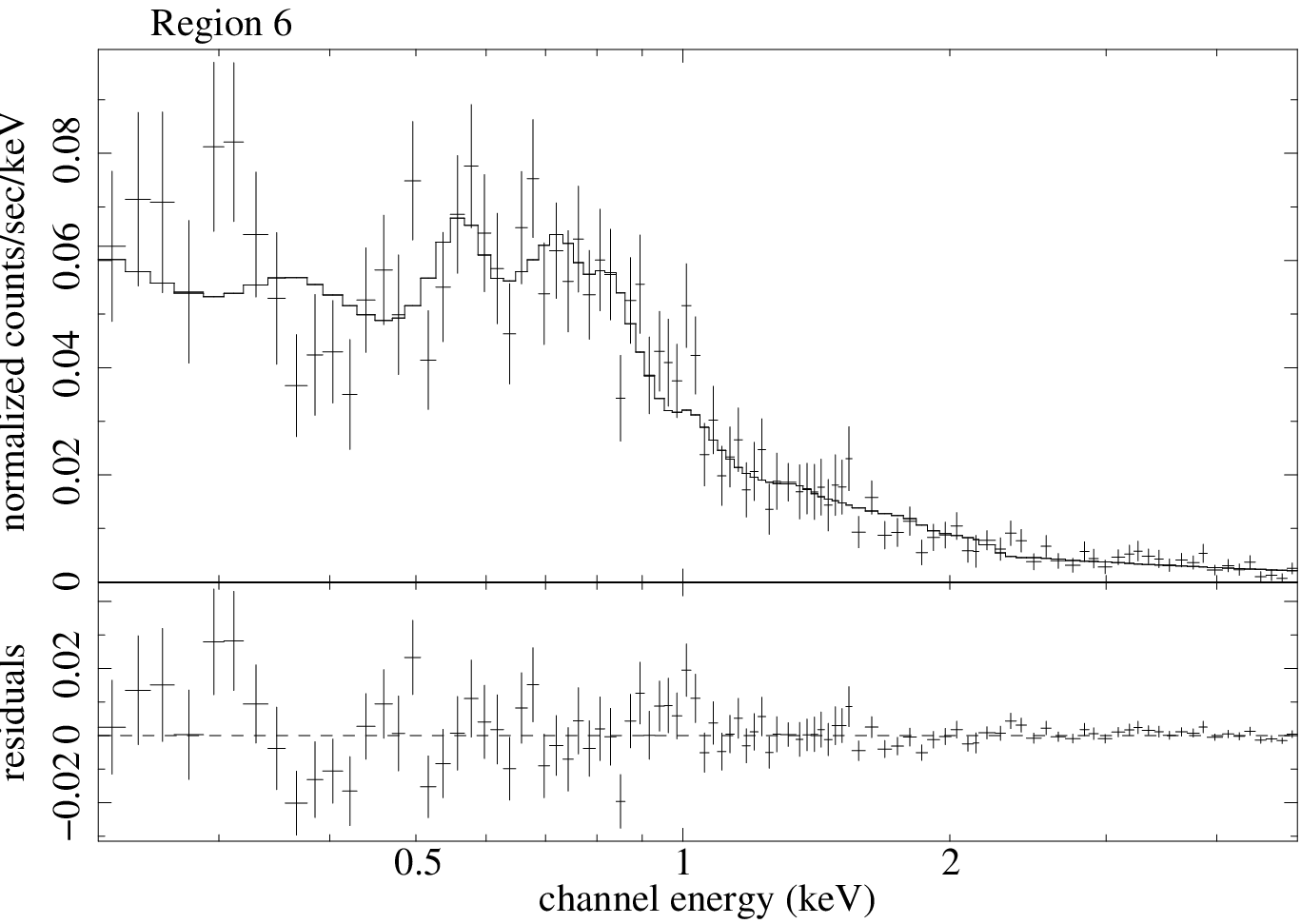}}
\resizebox{5cm}{!}{\includegraphics[angle=-90]{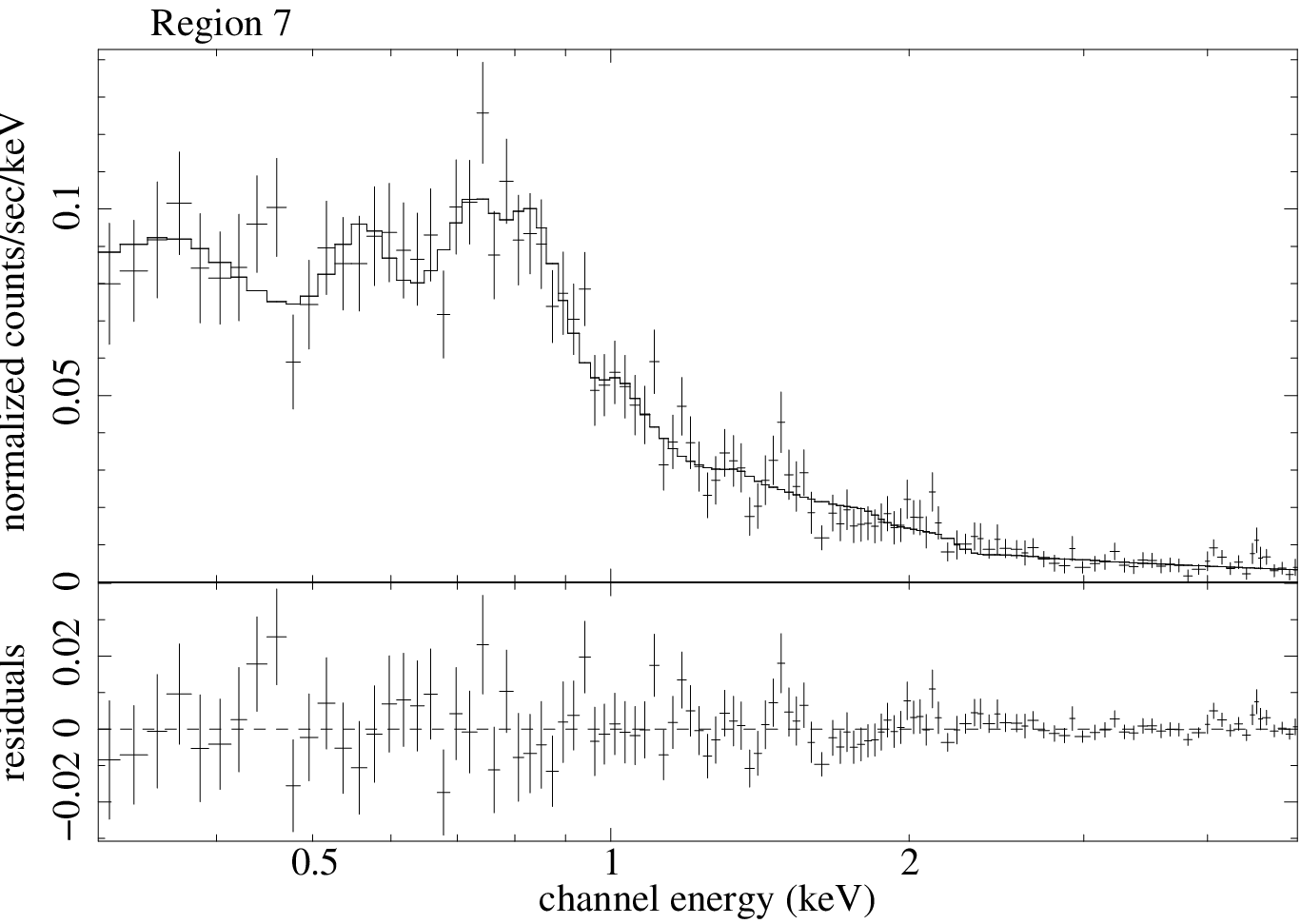}}
\caption{Model fits to the regions of NGC\,4569. See Tables~\ref{4569xtab} and
\ref{4569xf}.}
\label{4569mod}
\end{figure*}

Chy\.zy et al.~(\cite{chyzy4569}) discovered impressive radio lobes emerging from the disk of NGC\,4569, which is unusual for a normal
spiral galaxy. The authors argue that the lobes are a result of a strong past starburst phase.
We performed an analysis of our X-ray observations to search for clues that could confirm the proposed scenario of the origin and
evolution of the radio lobes: the extended X-ray emission from NGC\,4569 that we found suggests that there has been enhanced star formation resulting in hot gas
outflowing from the disk plane.
Extended soft X-ray emission from the galaxy (Fig.~\ref{4569xfig}) was modelled by
a two-temperature thermal plasma with a contribution from background point sources absorbed by the galactic foreground.
A two-temperature model seems to be a good approach in the case of studying outflows of the hot gas from the galactic disk, as it is the simplest 
approximation of the expected multi-temperature plasma, when the outflow gas mixes with the galactic halo. Adding more plasma components would require much more 
sensitive data. 
We constructed spectra for the regions presented in Fig.~\ref{4569xreg}, so that the emission from each region is used only once, e.g. the spectrum for region 6 does not include 
the emission from regions 2 and 4.
The cold and hot component in the galactic nucleus (region 1) have the temperatures of 
0.19$^{+0.07}_{-0.04}$~keV and 0.64$\pm$0.03~keV, respectively, with five times higher flux in the total energy band coming from the hotter component. 
We found only weak emission above 2~keV from this region, therefore the existence of a strong, active central source 
that might have produced the observed radio lobes is unlikely. 
Nevertheless, a weak nuclear X-ray source was identified in the high resolution Chandra data (Grier et al.~\cite{grier11}).
The galactic disk is characterized by two gas components of 0.14$^{+0.06}_{-0.04}$~keV and 0.46$\pm$0.07~keV. Here, the contribution from the hotter component is only roughly two times 
higher than that from the colder one. It is interesting to compare the temperatures of the hot gas in regions of visible H$\alpha$ outflow structures (regions 3 and 4) 
with those from areas of radio lobes on both 
sides of the disk. The western side shows a striking corespondence (compare regions 3 and 7 in Fig.~\ref{4569xreg} and Table~\ref{4569xtab}) of both cold and hot X-ray temperature components.
The eastern side, on the other hand, shows for the area of H$\alpha$ outflows much higher temperatures for both gas components than compared to the area of the radio lobe (regions 4 and 6 
in Fig.~\ref{4569xreg} and Table~\ref{4569xtab}). We discuss possible explanations for these discrepancies in Sect.~\ref{4569dis}.
In the area of the polarized radio ridge (region 5), we found gas components of 0.22$\pm$0.10~keV and 0.62$^{+0.24}_{-0.17}$~keV, with the hotter component having 
a temperature that is higher than for other regions (except for the hot component in region 4 - see Sect.~\ref{4569dis} for discussion). This increase is, however, only nominal, as it agrees 
within the uncertainties with the other temperature values of the hot components. 
The spectra for all regions together with the best-fit
models are presented in Fig.~\ref{4569mod}. The fit parameters and derived fluxes are presented in Tables \ref{4569xtab} and \ref{4569xf}. A thorough discussion of the fits
is presented in Sect.~\ref{4569dis}.

\begin{table}[ht]
\caption{\label{4569xtab}Model fit parameters of selected regions in NGC\,4569.}
\centering
\begin{tabular}{cccccc}
\hline\hline
Reg. & kT$_1$              & kT$_2$                 & Photon                  & Net & Red.    \\%& nhp$^{\ast}$ \\
\vspace{5pt} no.  & [keV]  & [keV]                  & Index                   & counts\tablefootmark{a} & $\chi^2$\\%&              \\
\hline
\vspace{5pt}
1 & 0.19$^{+0.07}_{-0.04}$ & 0.64$\pm$0.03          & 1.73$^{+0.09}_{-0.08}$  & 4238 & 1.4 \\%& 0.022 \\
%\vspace{5pt}
%  & 0.62$^{+0.02}_{-0.03}$ & --                            & 1.88$\pm$0.07           & 1.5 \\%& 0.006 \\
\vspace{5pt}
2 & 0.14$^{+0.06}_{-0.04}$ & 0.46$\pm$0.07          & 1.35$^{+0.39}_{-0.45}$  & 1071 & 0.9 \\%& 0.602\\
% 2 & 0.39$^{+0.06}_{-0.05}$ & --                           & 1.84$\pm$0.21           & 0.9 \\%& 0.695 \\
\vspace{5pt}
3 & 0.11$^{+0.01}_{-0.02}$ & 0.47$^{+0.08}_{-0.12}$ & 1.22$^{+0.25}_{-0.20}$  & 1141 & 1   \\%& 0.530\\
% 3 & 0.42$^{+0.12}_{-0.09}$ & --                           & 2.19$^{+0.15}_{-0.17}$  & 0.9 \\%& 0.638 \\
\vspace{5pt}
4 & 0.29$^{+0.09}_{-0.06}$ & 0.83$^{+0.63}_{-0.29}$ & 2.28$^{+0.32}_{-0.31}$  & 815  & 1.2 \\%& 0.272\\
% 4 & 0.37$^{+0.09}_{-0.06}$ & --                           & 2.23$^{+0.24}_{-0.29}$  & 1.3 \\%& 0.138 \\
\vspace{5pt}
5 & 0.22$\pm$0.10          & 0.62$^{+0.24}_{-0.17}$ & 1.41$^{+0.24}_{-0.26}$  & 1054 & 0.9 \\%& 0.728 \\
%\vspace{5pt}
% & 0.32$^{+0.16}_{-0.05}$ & --                     & 1.54$\pm$0.21           & 0.8 \\%& 0.782 \\
\vspace{5pt}
6 & 0.11$\pm$0.02          & 0.39$^{+0.11}_{-0.08}$ & 1.31$^{+0.23}_{-0.17}$  & 3552 & 1.1 \\%& 0.222 \\
\vspace{5pt}
7 & 0.10$^{+0.01}_{-0.02}$ & 0.47$^{+0.06}_{-0.08}$ & 1.32$^{+0.14}_{-0.15}$  & 5590 & 1.0 \\%& 0.447 \\
\hline
\end{tabular}
\tablefoot{
\tablefoottext{a}{Total net counts in the 0.2-12 keV band.}
}
\end{table}

\begin{table}[ht]
\caption{\label{4569xf}Total (0.2 - 12 keV) and soft (0.2 - 1 keV) unabsorbed fluxes in 10$^{-14}$erg\,cm$^{-2}$s$^{-1}$ for the modelled regions in NGC\,4569.}
\centering
\begin{tabular}{ccccc}
\vspace{5pt} Reg. no. & mekal 1  & mekal 2  & powerlaw & total    \\
\hline\hline
\vspace{5pt}
1 (total)             & 1.2$^{+0.6}_{-0.7}$ & 6$\pm$1 & 17$^{+2}_{-3}$ & 24$\pm$4 \\
1 (soft)	      & 1.1$\pm$0.7 & 4.2$^{+0.6}_{-1.1}$ & 4.5$^{+0.9}_{-0.8}$ & 9.8$^{+2.2}_{-2.6}$\\  
%\vspace{5pt}
%                     & 5.55$^{+0.56}_{-0.42}$ & --   & 16.72$^{+2.31}_{-1.79}$ & 22.28$^{+2.86}_{-2.21}$ \\
\vspace{5pt}
2 (total)             & 0.7$^{+1.4}_{-0.6}$ & 1.6$^{+0.5}_{-0.4}$ & 5$^{+8}_{-3}$ & 7$^{+10}_{-3}$ \\
2 (soft)	      & 0.7$^{+1.3}_{-0.6}$ & 1.3$^{+0.3}_{-0.2}$ & 0.7$^{+0.6}_{-0.3}$ & 2.7$^{+2.2}_{-1.2}$\\
% 2                   & 1.53$^{+0.41}_{-0.37}$ & --   & 4.18$^{+1.50}_{-0.99}$ & 5.71$^{+1.90}_{-1.35}$ \\
\vspace{5pt}
3 (total)             & 1.8$^{+0.6}_{-0.9}$ & 1.5$\pm$0.4 & 8$^{+6}_{-4}$ & 12$^{+7}_{-5}$ \\
3 (soft)	      & 1.8$^{+0.9}_{-0.7}$ & 1.2$^{+0.2}_{-0.3}$ & 0.8$^{+0.6}_{-0.3}$ & 3.8$^{+1.7}_{-1.3}$\\ 
%3                    & 1.02$^{+0.40}_{-0.36}$ & --   & 5.64$^{+1.01}_{-0.70}$ & 6.65$^{+1.42}_{-1.05}$ \\
\vspace{5pt}
4 (total)             & 0.7$^{+0.4}_{-0.5}$ & 0.4$^{+0.4}_{-0.3}$ & 2$\pm$1 & 3$\pm$2 \\
4 (soft)	      & 0.6$^{+0.3}_{-0.5}$ & 0.3$^{+0.3}_{-0.2}$ 1.2$^{+0.7}_{-0.5}$ & 2.1$\pm$1.3\\ 
%4                    & 0.82$^{+0.38}_{-0.31}$ & --   & 2.69$^{+0.64}_{-0.43}$ & 3.51$^{+1.01}_{-0.73}$ \\
\vspace{5pt}
5 (total)             & 1$^\pm$0.7 & 0.9$^{+0.5}_{-0.8}$ & 11$^{+7}_{-4}$ & 13$^{+8}_{-5}$ \\
5 (soft)	      & 1$^{+0.5}_{-0.6}$ & 0.7$^{+0.5}_{-0.6}$ & 1.6$^{+0.6}_{-0.5}$ & 3.2$^{+1.6}_{-1.7}$\\  
%\vspace{5pt}
%                     & 1.37$^{+0.61}_{-0.66}$ & --   & 10.34$^{+5.06}_{-2.75}$ & 11.71$^{+5.67}_{-3.41}$ \\
\vspace{5pt}
6 (total)             & 3.4$^{+5.2}_{-1.4}$ & 3$\pm$1 & 25$^{+11}_{-7}$ & 31$^{+18}_{-9}$ \\
6 (soft)	      & 3.4$^{+2.5}_{-1.9}$ & 2.2$^{+0.7}_{-0.6}$ & 3.2$^{+0.9}_{-0.7}$ & 8.7$^{+4.1}_{-3.1}$\\ 
\vspace{5pt}
7 (total)             & 7.6$^{+6.9}_{-2.2}$ & 5$\pm$1 & 49$^{+15}_{-9}$ & 61$^{+22}_{-12}$ \\
7 (soft)	      & 7.6$^{+4.4}_{-3.8}$ & 4.2$^{+0.5}_{-0.7}$ & 6.4$^{+1.6}_{-1.3}$ & 18.2$^{+6.5}_{-5.8}$\\ 
\hline
\end{tabular}
\end{table}

\subsection{NGC\,2276}
\label{2276spec}

\begin{figure}[ht]
\resizebox{\hsize}{!}{\includegraphics{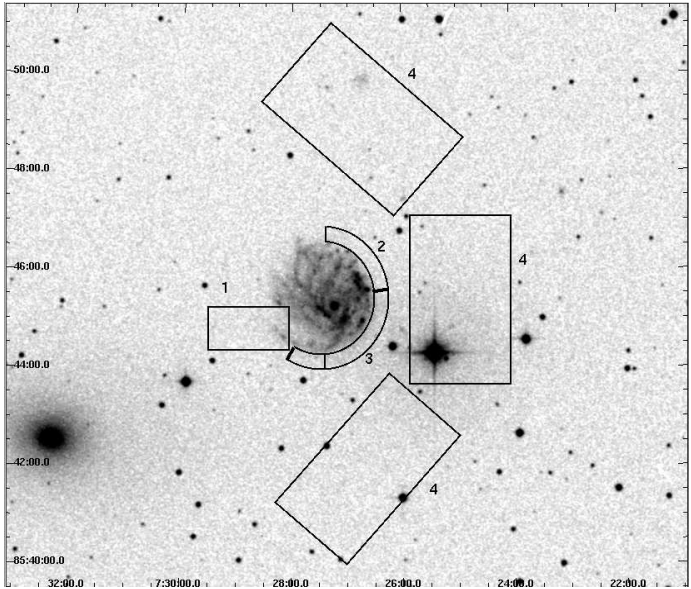}}
\caption{Regions of NGC\,2276 for which the spectra were acquired overlaid on the DSS blue image. All background sources were excised before extracting each spectrum.
The ends of the regions 2 and 3 are delineated with thin and thick lines, respectively. Three independent areas make up region 4.}
\label{2276xreg}
\end{figure}

As mentioned above, for NGC\,2276 we repeated part of the analysis performed by Rasmussen et al.~(\cite{rasmus}). 
The regions from which we extracted spectra are shown in Fig.~\ref{2276xreg}. Regions 2 (a bow-shock region from the analysis of Rasmussen et al.~\cite{rasmus}) and 3 (polarized ridge), being parts
of the same annular region, are indicated by thin and thick lines, respectively. For all regions, we followed Rasmussen et al.~(\cite{rasmus}) and fitted
a single thermal plasma model, fixing the abundance to 0.17 solar. For the region of a hot gaseous tail (region 1 in Fig.~\ref{2276xreg}), we obtained a temperature 
of kT = 0.96$^{+0.18}_{-0.20}$ keV (Table~\ref{2276xtab}). Our analysis of the XMM-Newton data for region 2 resulted in a temperature of 0.79$^{+0.18}_{-0.20}$ keV, 
which is the same as for the region of the hot IGM 
(region 4), 0.77$\pm$0.05 keV. 
However, owing to the lower spatial resolution of the XMM-Newton data, we needed to account for the possible influence of the point sources along 
the western side of the galactic disk, which we took care of by adding a power-law component to the model. 
This, however, has no impact on the temperature of the hot gas derived from the fitted model. 

\begin{figure*}[ht]
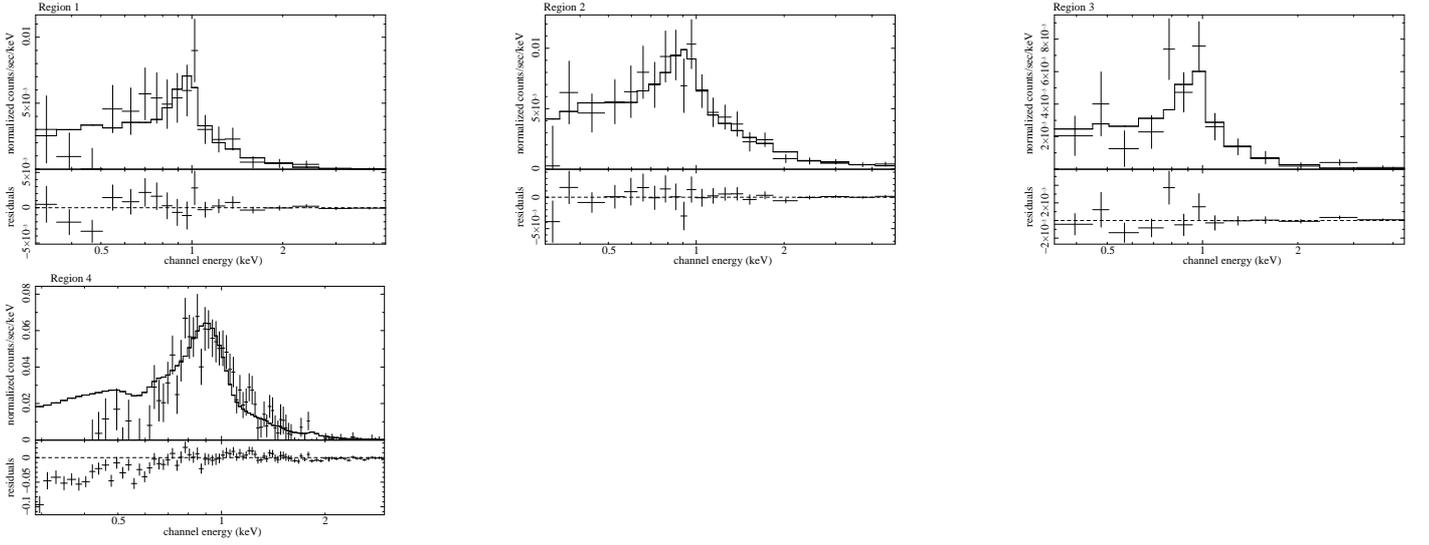

\resizebox{5cm}{!}{\includegraphics[angle=-90]{2276_region_1.ps}}
\resizebox{5cm}{!}{\includegraphics[angle=-90]{2276_region_2.ps}}
\resizebox{5cm}{!}{\includegraphics[angle=-90]{2276_region_3.ps}}
\resizebox{5cm}{!}{\includegraphics[angle=-90]{2276_region_4_all.ps}}
\caption{Model fits to the regions of NGC\,2276. See Tables~\ref{2276xtab} and
\ref{2276xf}. For the region 4, following Rasmussen et al.~(\cite{rasmus}), we exclude data for energies below 0.7~keV from the fit.}
\label{2276mod}
\end{figure*}

To investigate in more detail the region of the possible compression,
we produced a separate spectrum of the hot gas outside the polarized ridge (see Fig.~4b in Hummel \& Beck~\cite{humbeck} and region 3 in Fig.~\ref{2276xreg} in this work), from which we 
derived the hot gas temperature of 0.88$^{+0.16}_{-0.14}$ keV. 
Although this temperature agrees within errors with the temperature in the bow-shock region (region 2), 
%we fitted the spectrum of this region again, this time with a two temperature model, to check for an indication 
%of the hotter gas in the vicinity of a polarized ridge. We obtained temperatures of 0.34 and 0.91 keV with two times higher flux in the total energy band coming from the hotter component. 
%Unfortunately, the fit parameters were not constrained and no error derivation was possible. Nevertheless, 
this might be indicative of heated gas, mainly in the region of the radio polarized ridge. For the regions of hot IGM (region 4), following Rasmussen et al.~(\cite{rasmus}) we used the band 0.7 - 5 keV.
This is needed to avoid an overestimation of the background level in the blank-sky files at low energies owing to the relatively large $n_H$ of the source data (cf. Table~\ref{xobs} and 
Fig.~\ref{2276mod}). 
For all regions, the spectra together with the best-fit 
models are presented in Fig.~\ref{2276mod}. The fits parameters and derived fluxes are presented in Tables \ref{2276xtab} and \ref{2276xf}. A thorough discussion of the fits 
is presented in Sect.~\ref{2300dis}.

\begin{table}[ht]
\caption{\label{2276xtab}Model fit parameters of selected regions in NGC\,2276.}
\centering
\begin{tabular}{cccccc}
\hline\hline
Reg. & kT 			   & Photon  			& Net & Red.  \\%& nhp$^{\ast}$ \\
\vspace{5pt} no.  & [keV]  	   & Index         		& counts$^{\ast}$ & $\chi^2$ \\%& \\
\hline
\vspace{5pt}
1 & 0.96$^{+0.18}_{-0.20}$ 	   & --  			& 165 & 0.8 \\%& 0.715 \\
\vspace{5pt}
2 & 0.79$^{+0.18}_{-0.20}$ 	   & 1.32$^{+0.35}_{-0.22}$  	& 406 & 0.5 \\%& 0.940 \\
\vspace{5pt}
3 & 0.88$^{+0.16}_{-0.14}$ 	   & --  			& 168 & 1.1 \\%& 0.354 \\
\vspace{5pt}
4${^\ast}{^\ast}$ & 0.77$\pm$0.05  & -- 			& 850 & 1.2 \\%& 0.027 \\
\hline
\end{tabular}
\tablefoot{
\tablefoottext{a}{Total net counts in the 0.3-5 keV band except for region 4 (see below).}
\tablefoottext{b}{in the band 0.7 - 5 keV (see text).}
}
\end{table}

\begin{table}[ht]
\caption{\label{2276xf}Total (0.3 - 5 keV) and soft (0.2 - 1 keV) unabsorbed fluxes in 10$^{-14}$erg\,cm$^{-2}$s$^{-1}$ for modelled regions in NGC\,2276.}
\centering
\begin{tabular}{ccccc}
\vspace{5pt} Reg. no. & mekal & powerlaw  & total \\
\hline\hline
\vspace{5pt}
1 (total) & 1.0$\pm$0.2 &  -- 		    & 1$\pm$0.2 \\
1 (soft) & 0.6$\pm$0.2 & -- & 0.6$\pm$0.2\\
\vspace{5pt}
2 (total) & 1.4$^{+0.6}_{-0.4}$ & 2.8$\pm$2 & 4.2$^{+2.6}_{-2.2}$ \\
2 (soft) & 0.9$^{+0.4}_{-0.2}$ & 0.6$^{+0.5}_{-0.4}$ & 1.5$^{+0.9}_{-0.6}$\\  
\vspace{5pt}
3 (total) & 1.1$\pm$0.2 & -- 		    & 1.1$\pm$0.2 \\
3 (soft) & 0.7$\pm$0.2 & -- & 0.7$\pm$0.2\\ 
\vspace{5pt}
4$^{\ast}$ (total) & 7.3$\pm$0.6 	   & --     & 7.3$\pm$0.6 \\
4$^{\ast}$ (soft) & 3.9$^{+0.6}_{-0.5}$ & -- & 3.9$^{+0.6}_{-0.5}$\\
\hline
\end{tabular}
\tablefoot{
\tablefoottext{a}{in the band 0.7 - 5 keV (see text).}
}
\end{table}

\section{Discussion}
\label{disc}

\subsection{Possible tidal interactions of NGC\,4254}
\label{4254dis}

NGC\,4254 was the target of the radio observations of Soida et al.~(\cite{soida96}), as well as the MHD simulations of Vollmer et al.~(\cite{vollmer05}). The latter authors suggested that 
this galaxy may be experiencing ongoing ram-pressure stripping. However, NGC\,4254 was thoroughly studied later in the radio domain with higher spatial resolution and 
sensitivity by Chy\.zy~(\cite{chyzy4254b}). The author argued that the distorted
optical morphology and a region of bright polarized intensity visible in this galaxy are not caused by compression effects but
result from tidal interactions with another Virgo cluster member (probably VIRGO~HI~21, Minchin et al.~\cite{minchin}).
The polarized ridge would thus be produced
by pulling the spiral arm out of the galaxy (northwards) and stretching the magnetic field in the southern
part of the disk, where the ridge is visible (Chy\.zy~\cite{chyzy4254b}).

In the eastern arm of NGC\,4254 (region 1 in Fig.~\ref{4254xreg}), 
the temperature of the hot gas seems to be slightly higher than in the other arms (regions 2 and 3), though still agrees within the errors. 
In the southern part of the galaxy where 
a polarized radio ridge is visible (region 4), a single-temperature fit yields a value similar to that for arms 2 and 3. However a two-temperature fit 
suggests that the hot gas in this area is not in thermal equilibrium and that one can trace its components at different temperatures. The hotter one is fitted with a temperature of 
0.64$^{+0.12}_{-0.11}$~keV, which is higher than in other areas of the galaxy. This increase in temperature can be explained, as also in the case of a higher temperature in the eastern arm, 
by the existence of bright H$\alpha$ clumps (see Fig.~\ref{4254xreg}). 
They are regions of high star formations, which heat the surrounding ISM more intensively via supernovae explosions.
As we mentioned in Sect.~\ref{4254spec}, a detection of a hotter component in the model fit for the region 4 could be also owing to photon statistics. Then, we could expect a similar 
component in the other model fits, provided that the data are sensitive enough. Nevertheless, since NGC\,4254 is a galaxy rapidly forming stars, the hotter component could be 
a result of enhanced star forming activity.
Therefore, it is unlikely that the increase in the temperature of the hot gas in region 4 is produced by a shock compression, which would explain the existence of a polarized radio 
ridge. Especially, that the outer part of the polarized ridge area (region 7) is even cooler. The temperature of the region 7 is only 0.14$\pm$0.04 keV, possibly owing 
to a large \ion{H}{i} cloud infalling into NGC\,4254 from the south 
(Phookun et al.~\cite{phookun}, Chy\.zy et al.~\cite{chyzy4254}), which causes a decrease in the temperature of the hot gas by thermal conduction.
Given this argumentation, we conclude that our spectral analysis of the X-ray data support the scenario for the origin of the polarized radio ridge given by 
Chy\.zy~(\cite{chyzy4254b}): it is most likely caused by tidal interactions (with the companion galaxy), 
which created shearing forces that eventually led to the formation of a polarized radio feature. 

\subsection{Past starburst in NGC\,4569}
\label{4569dis}

The extended, diffuse radio lobes detected by Chy\.zy et al.~(\cite{chyzy4569}) were explained by the authors as having been formed during a past starburst phase.
This galaxy is bluer in B-V colour by 0$\fm$11 than a typical
galaxy of the same type and at the same time redder in U-B colour by 0$\fm$17, which the traces overabundance of medium-aged A-stars and some deficit of youngest ones. 
This means that we observe this galaxy
after the end of the starburst phase,  whose effects are now diminishing. This scenario would also explain the significant strength of the 
magnetic field of NGC\,4569 (We\.zgowiec~\cite{phd}, Chy\.zy et al.~in prep.), which is quite surprising for an anaemic galaxy. Our analysis of the X-ray emission 
from the nuclear region also confirms that the radio lobes are most likely products of a past galactic starburst phase, as in the X-ray data
we find no spectral signs of an active galactic nucleus in its centre, that could produce the observed radio lobes/jets. The modelled power-law component 
of the fit to the central region can suggest the existence of only a low X-ray luminosity nuclear component.

In We\.zgowiec et al.~(\cite{wezgowiec11}), an extended hot-gas halo was presented that shows a Mach cone-like geometry suggesting that the galaxy has undergone 
interactions with the ICM. This was, however, an 
analysis of the global hot-gas envelopes around the spiral galaxies of the Virgo cluster. Here, we analyse the features found within such a gaseous envelope in detail. Therefore, 
the low surface-brightness cone-like structure around NGC\,4569 shown in the 1' map in We\.zgowiec et al.~(\cite{wezgowiec11}) 
is not visible when presented in a map with a higher resolution (Fig.~\ref{4569xfig}).
The detection of a Mach cone filled with hot gas could explain the symmetry of the observed radio lobes. The shock front might have reduced the ram pressure exerted on the 
hot gas inside the Mach cone. The higher internal pressure in the radio lobes would then allow them to freely expand into this gas, 
still within the Mach cone (see We\.zgowiec et al.~(\cite{wezgowiec11}) for more details). The configuration of the Mach cone structure also suggests 
that the galaxy is moving in the northeast direction. Here, we examine the X-ray emission from hot gas features
coincident with H$\alpha$ outflows and extended radio lobes, including the region of the polarized radio ridge (see Sect.~\ref{4569spec}). 

As mentioned in Sect.~\ref{4569spec}, the hot gas in the regions of H$\alpha$ outflows significantly differs between both sides of the disk. 
Assuming that the orientation of the disk derived from the dust lanes (Bomans et al., in prep.) is correct, 
the first step in obtaining a consistent explanation is to examine the GALEX UV image of NGC 4569 (Boselli et al.~\cite{boselli06}). The star formation 
is asymmetric between the eastern and the western parts of the disk. On the eastern side, which is closer to us, the UV emission 
is less extended than in the western side, where an additional arm contributes significantly. This arm, which is also visible in H$\alpha$, 
is not connected to the main disk but appears to be out of the disk. Such a geometry would imply that there are more numerous and denser gas clouds 
in both the thick disk and lower halo in the west than in the east. Therefore, a galactic outflow can freely expand out of the disk towards 
the eastern halo, while the flow to the west interacts with clouds and the anomalous arm. As shown in Bomans et al.~(in prep.), the H$\alpha$ kinematics
even imply that there has been an interaction with additional clouds at larger distances. The result of these different expansion conditions would be a hotter 
lower-density gas in the outflow towards the east, since the density and therefore the cooling rate is lower. The X-ray gas would probably 
also move out of equilibrium, with fast adiabatic cooling, but owing to the longer recombination time it maintains the line pattern for a hotter gas, which 
we measure by analysing the X-ray line spectrum. The X-ray emission indeed shows a higher temperature on this side of 
the disk (region 4 in Fig.~\ref{4569xreg} and Table~\ref{4569xtab}). 
%This line of arguments naturally implies that the surface brightness of both the X-ray emitting gas, but especially the H$\alpha$ emitting 
%gas (the denser, cooler gas, from the compressed walls of the outflow and entrenched or overrun dense clouds) would be of much lower surface 
%brightness in the east.  
%Exactly these clouds are present in the western side above the disk, e.g. as a part of the arm, where the H$\alpha$ images shows knots and filaments. 
%The non-detection of similar, but fainter features towards the east would therefore be 
%a problem of the surface brightness limit of the image and we predict to detect a more symmetric H$\alpha$ structure on a significantly more 
%sensitive image. 
One potentially puzzling point with this interpretation is the X-ray 
temperature of the gas further out in the halo, where the temperatures of both the eastern and western sides of the halo agree within the errors (see Table~\ref{4569xtab}). 
Nevertheless, we have to keep in mind that the outflow does not expand into an ICM, but into a Mach cone (We\.zgowiec et al.~\cite{wezgowiec11}), 
which effectively confines and isolates the region, while also contributing to its energetics. This may result in a semi-equilibrium condition between heating and cooling, 
at the temperatures measured in the outer halo. These two temperatures (0.1 and 0.5 keV) are actually close to a local minimum of the cooling curve 
and in a region of rapidly falling cooling-rate (e.g. Sutherland \& Dopita~\cite{sutherland93}) for solar metallicity plasma.

It is also interesting to consider a region corresponding to a polarized radio ridge, southwest of the galaxy centre (region 5 in Fig.~\ref{4569xreg}). 
We argue that this feature is caused by compression effects rather than a tidal influence. The extended radio lobes remain symmetric and show no signs of external distortions. 
The polarized radio ridge would then be produced by the local compression of the gas that made the magnetic field more regular. Such a compression would be consistent with 
an increase in the temperature of the hot gas at this position, which is a sign of shock heating.
The hotter component in this region is as hot as 0.62$^{+0.24}_{-0.17}$~keV. We suggest that this increase in the temperature, when compared to the hot component of region 7, 
though not statistically significant, might be evindence of additional heating taking place in this area. This would agree with the explanation of Chy\.zy et 
al.~(\cite{chyzy4569}), who argue that the polarized radio ridge results from the shocking of previously stripped \ion{H}{i} material (falling back onto the galaxy from the southwest)
by the galactic wind. 
The same effect is also apparent in the kinematics of the H$\alpha$.
The wind flowing with more than 100 km/s projected radial 
velocity is decelerated to rest velocity near the top of our 
region 5 (Bomans et al., in prep).
Assuming a temperature of the hot gas in the radio lobes to be of the order of 0.3~keV (we base our assumption on the resultant temperatures of the radio lobes - see Table~\ref{4569xtab}), we estimate 
the local sound speed to be 260 km/s. Using Eq.~2, we calculate that the temperature of 0.62~keV requires a shock moving at the velocity of around 540~km/s. 
This velocity is consistent with a scenario of a starburst phase and the formation of extended radio lobes from the expanding galactic wind into the hot ICM. The value of the inferred 
expansion velocity of the wind is high, but not inconsistent with the measured radial component of the flow (Bomans et al., in prep.).

\subsection{Mixed interactions in the NGC\,2300 group?}
\label{2300dis}

NGC\,2276 was reported by Rasmussen et al.~(\cite{rasmus}) to be a case of a supersonic motion causing 
ram pressure effects strong enough to form a bow shock at the leading edge of the galactic disk. 
The radio observations of Hummel \& Beck~(\cite{humbeck}) however seem to favour tidal interactions between NGC\,2276 and NGC\,2300 as a possible cause of the observed features. 

\begin{figure}[ht]
                        \resizebox{\hsize}{!}{\includegraphics{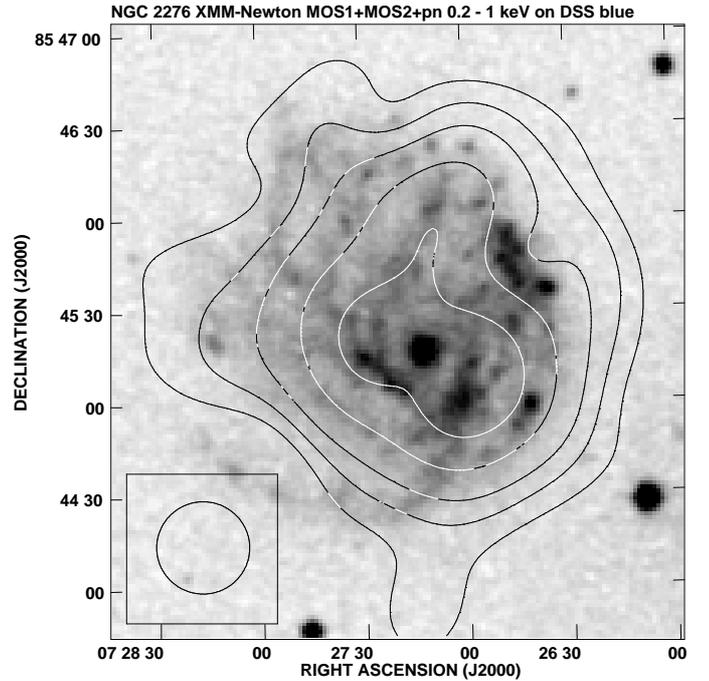}}
\caption{
Central 3$\arcmin$ of Fig.~\ref{2276xfig}. The resolution is 30$\arcsec$ and the contours are
13, 18, 25, 40, and 60 $\times$ rms. The greyscale is that of the DSS blue image. The smoothing scale is shown in the lower left corner of the figure.
}
\label{2276xfig_zoom}
        \end{figure}

A relatively uniform distribution of the emission in our data do not seem to be a resolution effect, 
as the emission gradient seen in the Chandra image (Rasmussen et al.~\cite{rasmus}) spans some 0$\farcm$5 and our
image has been smoothed to a resolution of 10'' (Fig.~\ref{2276xfig}). Therefore, if a strong gradient existed in our data, we would still be able to detect it.
One can however see that the hot gas has a somewhat asymmetrical distribution outside the disk of NGC\,2276 in the low resolution image (Fig.~\ref{2276xfig_zoom}).
Provided that the galaxy is moving to the west, this could be an argument for a subsonic ``plowing'' of the IGM gas by the galaxy.

Our analysis shows that the temperature
of the hot gas in the ''shock front'' (region 2 in Fig.~\ref{2276xreg}) is negligibly higher than that of the hot IGM (by 0.02~keV), 
therefore, given the uncertainties in both values we can conclude that no heating takes place and that the galaxy is moving subsonically. 
This agrees with the value of the radial velocity relative to the IGM provided in Table~\ref{objects}, even when we assume that the tangential component of the galactic velocity 
is comparable to the radial one, which would lead to an actual velocity of the order of 340~km/s, which would still be subsonic.  

Although our results agree within errors with those presented in Rasmussen et al.~(\cite{rasmus}), the spectral 
analysis of the XMM-Newton data allows more precise constraints to be made on the derived fit parameters. 
We note here that this can be caused by the instrument-dependent differences between the Chandra and XMM-Newton observations.
The latter is much more sensitive to extended diffuse soft X-ray emission owing to its higher effective area, 
which makes the ''IGM background'' in the close vicinity of the galaxy clearly detectable. 
This in turn leads to a shallower intensity gradient 
between the galaxy and the surrounding IGM. Similarly for the spectra, a lower contribution from the softest energy range and the overall lower sensitivity 
of the Chandra data can lead to less accurate measurements of the temperatures 
(see the comparison of XMM-Newton and Chandra in the XMM-Newton Users Handbook for details\footnote{http://xmm.esac.esa.int/external/xmm\_user\_support/\\documentation/uhb}). 
This could then result in the determination of a higher temperature by Rasmussen et al.~(\cite{rasmus}) for the region outside the western edge of NGC\,2276 disk, which is claimed to be 
a bow shock.

The region of the polarized radio ridge (region 3 in Fig.~\ref{2276xreg}), however, has a temperature of 0.88$^{+0.16}_{-0.14}$~keV, 
which is slightly higher than that of the IGM (0.77$\pm$0.05~keV). Although the difference is still within errors, we may see hints of an admixture of hotter gas. 
%(also suggested by a 
%two-temperature model fit mentioned in Sect.~\ref{2276spec}). 
For the local speed of sound of 420~km/s, this higher temperature corresponds to a velocity of 480~km/s. Assuming that the velocity of the galaxy is of the order of 340~km/s (see above), 
the required increase in the velocity by 140~km/s can be explained by the rotation of the galactic disk, when the rotational velocity of the leading arm is added to the galactic velocity.    
We would then observe a compression due to ram pressure, but only along the southwestern edge of the disk, where the polarized radio ridge is visible. The lack of visible signs 
of the hot gas compression in this area (Figs.~\ref{2276xfig} and \ref{2276xfig_zoom}) can be explained by the low sensitivity of the X-ray data, but also by 
the radio polarimetry being a very sensitive tracer of even weak gas compressions, that are not clearly visible in other domains (Urbanik~\cite{urbanik05}).
Nevertheless, since the aforementioned increase in the temperature of the hot gas is not statistically significant and given the overall shape of the galaxy, including the stretched southern spiral arm, 
we may also conclude that the enhancement of the magnetic field in the region of the polarized radio ridge was produced by the tidal interactions, 
as in the case of NGC\,4254 (see Sect.~\ref{4254dis} and Chy\.zy~\cite{chyzy4254b}).

\section{Summary and conclusions}
\label{cons}

We have presented our analysis of extended X-ray emission from the two Virgo cluster spiral galaxies NGC\,4254 and NGC\,4569, and 
from NGC\,2276 in the NGC\,2300 galaxy group. NGC\,2276 and NGC\,4254 
have fairly uniform distributions of extended soft X-ray emission that follows the distribution of bright \ion{H}{ii} regions. NGC\,4569 has an extended halo of hot gas that is associated with a 
past starburst phase. Each galaxy shows a polarized radio ridge that may be a result of either ram pressure or tidal interaction effects. We have shown that 
the spectral analysis of the soft X-ray emission from these regions helps us to investigate how the polarized radio ridge was produced. We use the fact that in order to produce a compression 
that enhances and orders the magnetic field, a shock is needed that also heats the ISM. A lack of any evidence of such heating in the polarized radio ridge area 
implies shearing forces produced by tidal interactions that enhance and order the magnetic fields. After determining the temperatures of the different relevant areas we conclude as follows:

\begin{itemize}
\item[-] In NGC\,4254, we see no temperature increase in the polarized radio ridge (and even a decrease outside it), 
	 which suggests that it is produced by tidal interactions with a companion galaxy/\ion{H}{i} cloud.
\item[-] The area of the polarized radio ridge in the western radio lobe of NGC\,4569 
	 shows a slight increase in the hot gas temperature, which is an evidence of a compression caused by the galactic wind hitting previously stripped material.
\item[-] In NGC\,2276, the visible distortions are likely produced by a tidal interaction with NGC\,2300, although the observed polarized radio ridge may also be the result of the 
	combined effects of the rotation of the galaxy and its movement through the IGM, as suggested by an increase (though within errors) in the temperature of hot gas 
	in the area of the polarized radio ridge.
\end{itemize}

As we have shown above, our spectral analysis of the X-ray data proves to be useful in determining the origin of the perturbations observed at radio wavelengths. 
Therefore, it is important to study in the X-ray domain more perturbed group and cluster spiral galaxies, for which the nature of the observed perturbations is still under question.
The analysis of the relevant data will allow us to achieve a clearer understanding of the conditions under which galaxies interact with the cluster environment. 
It will also enable us to put more precise constraints on models of galactic evolution in a cluster or group environment. 

\begin{acknowledgements}
This work was supported by DLR Verbundforschung
"Extraterrestrische Physik" at Ruhr-University Bochum through
grant 50 OR 0801 and by the Polish Ministry of Science and Higher
Education, grants 2693/H03/2006/31 and 3033/B/H03/2008/35. We also thank the anonymous referee for comments that helped to improve this paper. 
We acknowledge the use of the HyperLeda database (http://leda.univ-lyon1.fr).
\end{acknowledgements}

\end{document}